\documentclass[final,onefignum,onetabnum]{siamltex1213}

\usepackage{framed,multirow,amsmath}
\usepackage{amssymb}
\usepackage{latexsym}
\usepackage{multicol}
\usepackage[mathscr]{eucal}
\usepackage{array}
\usepackage{hyperref}
\usepackage{amstext}
\usepackage{rotating}
\usepackage{subfig}
\usepackage{graphicx}
\usepackage{algorithm2e}
\usepackage{todonotes}

\newcommand{\be}{\begin{equation}}
\newcommand{\ee}{\end{equation}}
\newcommand{\bea}{\begin{eqnarray}}
\newcommand{\eea}{\end{eqnarray}}

\title{Stratifying High Dimensional Data Based on Proximity to the Convex Hull Boundary} 

\author{Lori Ziegelmeier\footnotemark[1]\ 
\and Michael Kirby\footnotemark[2]\ 
\and Chris Peterson\footnotemark[2]\ \thanks{This research was partially supported by NSF awards DMS-1228308 and DMS-1322508.
(\email{lziegel1@macalester.edu}). Questions, comments, or corrections
to this document may be directed to that email address.}}

\begin{document}
\maketitle
\slugger{mms}{xxxx}{xx}{x}{x--x}

\renewcommand{\thefootnote}{\fnsymbol{footnote}}
\footnotetext[1]{Macalester College, Saint Paul, MN 55104}
\footnotetext[2]{Colorado State University, Fort Collins, CO 80523}

\begin{abstract}
The convex hull of a set of points, $C$, serves to expose extremal properties of $C$ and can help identify elements in $C$ of high interest. For many problems, particularly in the presence of noise, the true vertex set (and facets) may be difficult to determine. One solution is to expand the list of high interest candidates to points lying near the boundary of the convex hull. We propose a quadratic program for the purpose of stratifying points in a data cloud based on proximity to the boundary of the convex hull.  For each data point, a quadratic program is solved to determine an associated weight vector.  We show that the weight vector encodes geometric information concerning the point's relationship to the boundary of the convex hull.  The computation of the weight vectors can be carried out in parallel, and for a fixed number of points and fixed neighborhood size, the overall computational complexity of the algorithm grows linearly with dimension. As a consequence, meaningful computations can be completed on reasonably large, high dimensional data sets.

\end{abstract}


\begin{AM} 52A20, 52A41, 90C20, 90C25 \end{AM}

\pagestyle{myheadings}
\thispagestyle{plain}
\markboth{L. Ziegelmeier, M. Kirby, and C. Peterson}{Stratifying Data in a Convex Hull}

\section{Introduction}

The convex hull of a set of points in Euclidean space is defined as the smallest convex set containing the points.  Convex hull computations have applications in a diverse collection of areas including number theory, combinatorics, algebraic geometry, pattern recognition, endmember detection, data visualization, path planning, and geographical information systems \cite{boardman, deberg,  lozano, O'Rourke, Schneider}.  This range of applications has helped spur the development of efficient algorithms and has contributed to its role as a fundamental tool in computational geometry. While there are multiple algorithms for computing the set of vertices for a convex hull, describing the full collection of facets is impractical for high dimensional data sets due to the rapid growth in the number of facets as a function of dimension. In many concrete applications, vertices and boundary points of the convex hull may carry crucial information, or clues, as to the intrinsic nature of the data.  For example, these points may capture pure or unmixed distinguishing features, or be potential optima for linear functionals on a data-defined feasible region.  The approach proposed here allows one to characterize sampled data in high-dimensions as vertices, boundary points, or points near the boundary, in addition to stratifying points in relation to proximity to the boundary of the convex hull.

Determining the convex hull of a finite set of points has a long history in computational geometry. The planar case is well-studied and includes methods such as Gift Wrapping, Quickhull, Graham's Algorithm, Divide and Conquer, and the Incremental Algorithm to name a few \cite{Barber, O'Rourke}.  Several of these algorithms may be extended to the higher dimensional case.  For instance, the Quickhull algorithm may be used, in theory, to determine all facets of a convex hull in $n$-dimensions.  However,  the rapid increase in the number of facets as a function of dimension leads to the problem becoming computationally intractable even for relatively small examples such as 20 dimensional data consisting of a few hundred points \cite{klee}.   For many problems of interest, particularly in the presence of noise, the true vertex set (and facets) may be difficult to determine and one should expand the list of extremal candidates to points lying near the boundary of the convex hull. Thus, for certain applications, determining the complete set of facets is less relevant than determining the near-the-boundary data. We propose a convex optimization problem and algorithm for coarsely ranking and stratifying data based on proximity  to the boundary of the convex hull which we call the \emph{Convex Hull Stratification Algorithm} (CHSA).  

\section{A First Example: Piles of Sand}
\label{firsteg}

\begin{figure}
\centering
\subfloat[]{\label{fig:Sand1Image}\includegraphics[width=.5\textwidth]{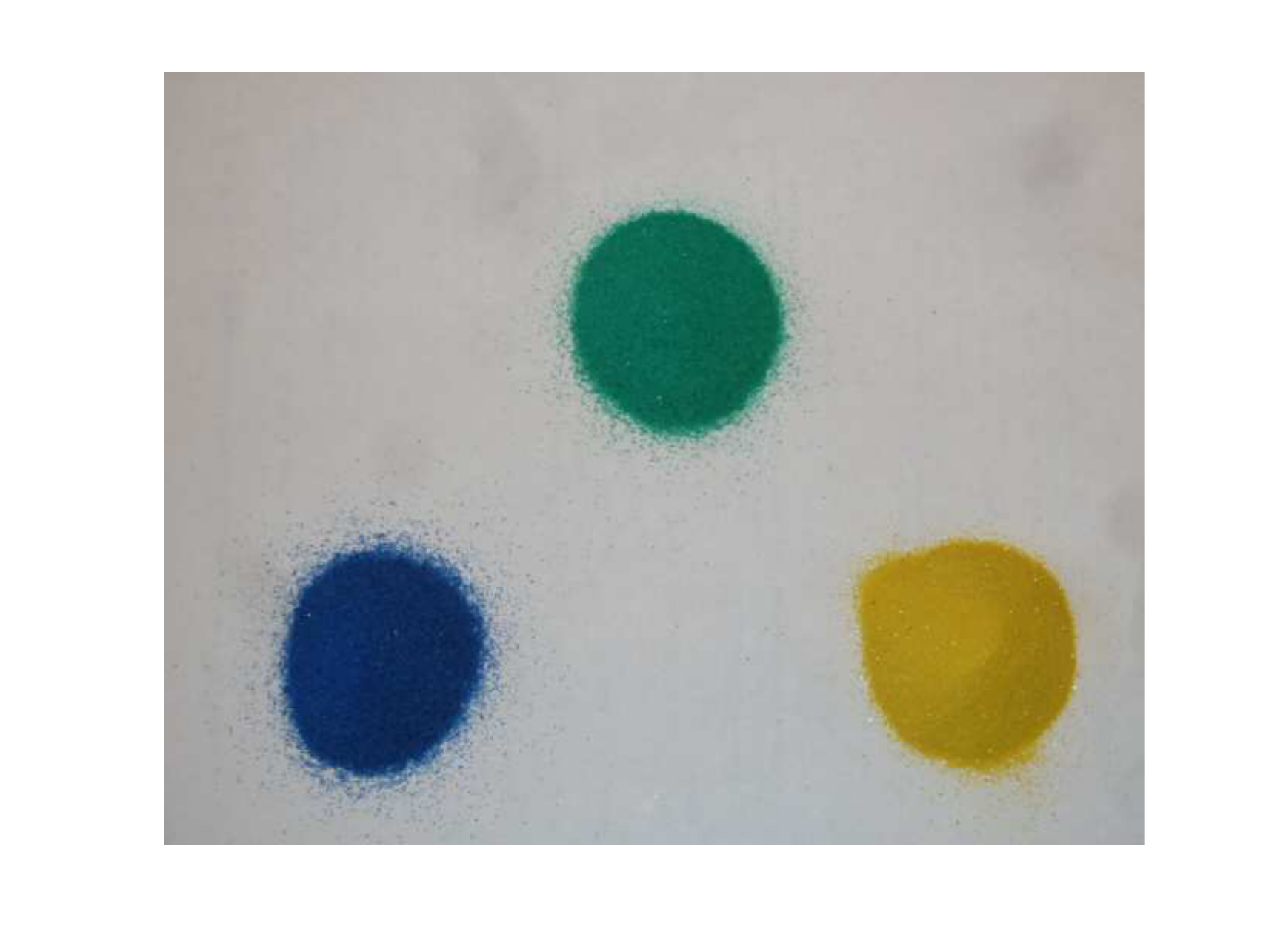}}
\subfloat[]{\label{fig:Sand2Image}\includegraphics[width=.5\textwidth]{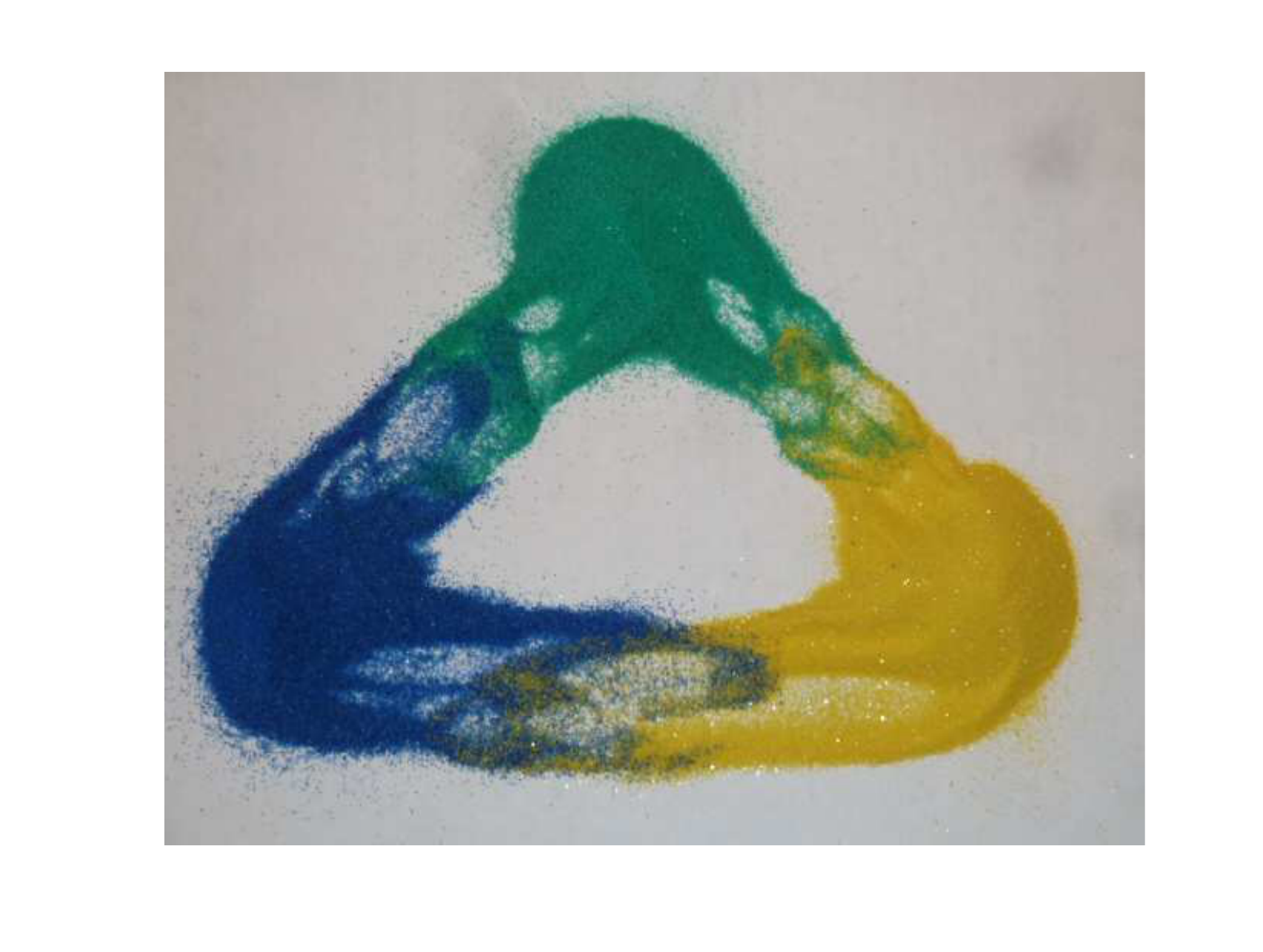}}\\
\subfloat[]{\label{fig:Sand3Image}\includegraphics[width=.5\textwidth]{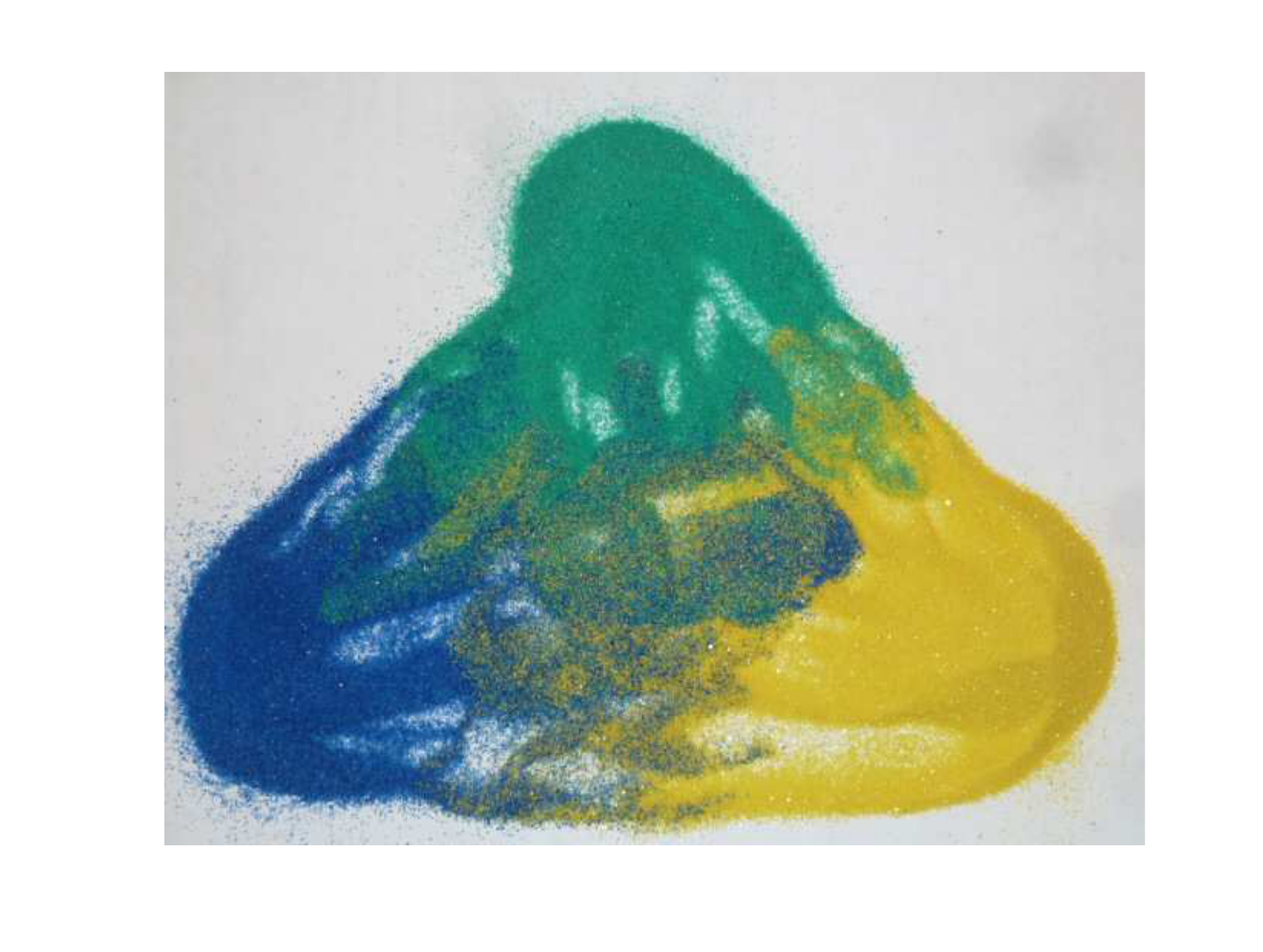}}
\subfloat[]{\label{fig:Sand4mage}\includegraphics[width=.5\textwidth]{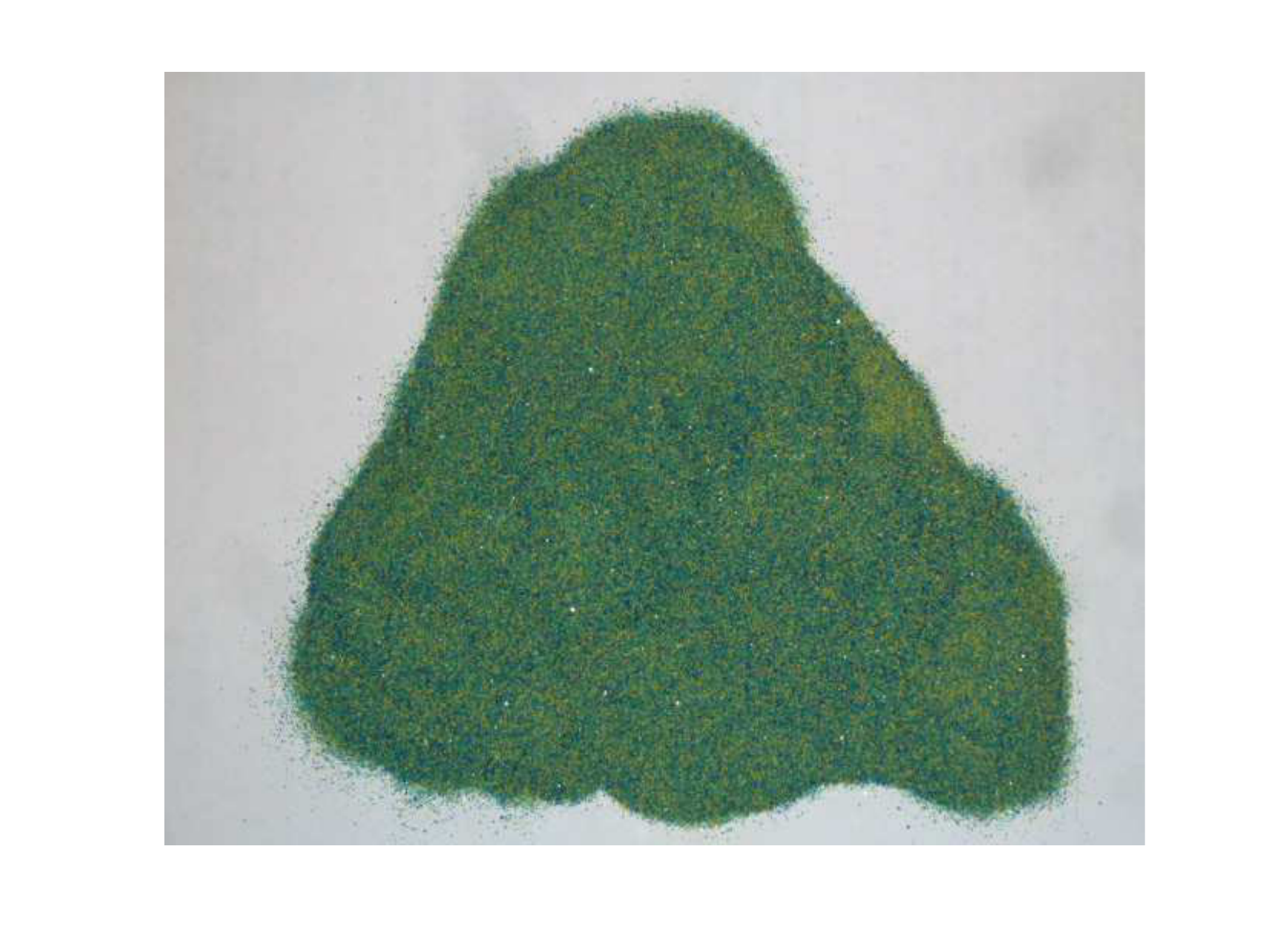}}
\caption{An experiment of sand being blended together at different levels of mixing.}
\label{fig:Sand}
\end{figure}

To motivate our approach, we present an example/experiment involving the identification of the ``purest pixels" within a digital image.  The experiment involves various levels of mixing of three different colors of sand. Images of the sand, lying on a sheet of white paper, are captured via a digital camera as shown in Fig. \ref{fig:Sand}.  In Fig.  \ref{fig:Sand}a, the sand lies unmixed in three piles, \emph{i.e.} with each pile consisting of a single, pure color.  In Fig.  \ref{fig:Sand}b, the three piles of sand are partially mixed, two piles at a time, leaving a portion of each monochromatic pile in its original unmixed state. In Fig.  \ref{fig:Sand}c, part of the sand from all three piles is mixed (still leaving a portion of each pile in its original unmixed state as well as a portion of sand involving a mixture of only two colors).
Finally, in Fig.  \ref{fig:Sand}d, the three piles of sand are thoroughly mixed together.  

A data set in $\mathbb R^3$ can be built from the color coordinates of individual pixels in each digital image.  This example will serve to illustrate how proximity to the convex hull of the data sets arising from each of the sand images helps to identify the purest pixels within each image. The next paragraph provides more detail on how one may consider each of these digital images as a data set in $\mathbb R^3$.

\begin{figure}[h]
 	\centering 
 	{\includegraphics[width=\textwidth]{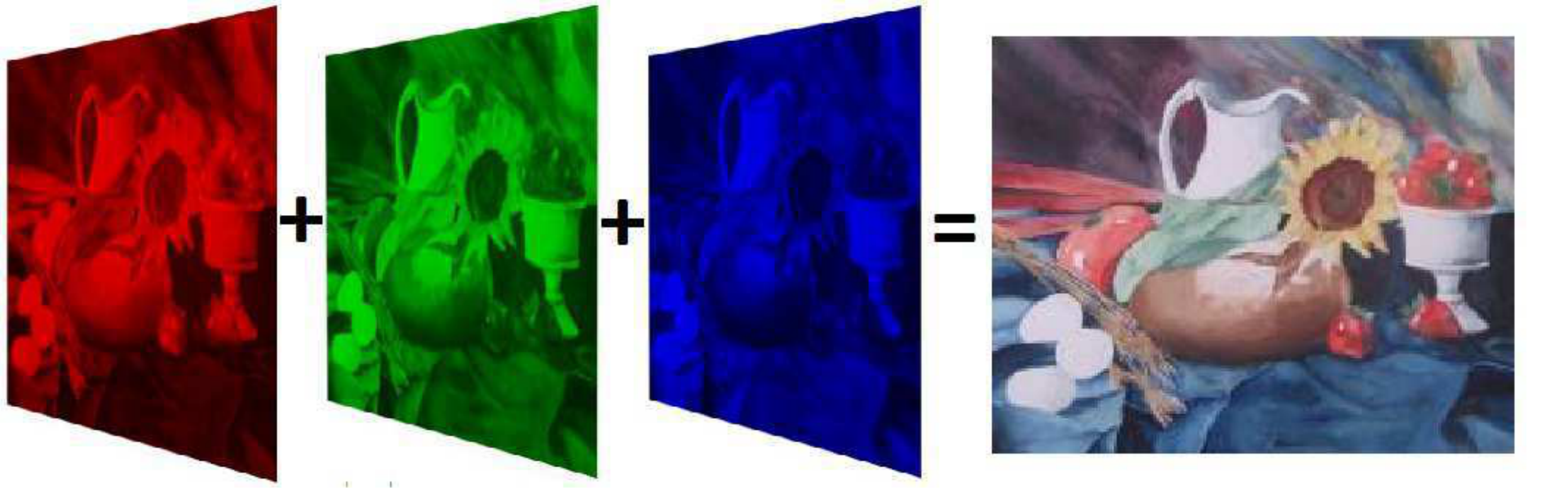}}
 	\caption{Illustration of the red, green, and blue sheets associated to pixels of an image.}
 	\label{fig:StillLifeSheets}
\end{figure}

Consider an illuminated scene and the resulting reflectance/scattering of the incoming light. From a given point of view, one can regard the points in the scene as parameterizing a family of {\it spectral reflectance curves}. The scene under a fixed illumination condition, as perceived by the human eye, can roughly be viewed as a map to $\mathbb{R}^3$ obtained by integrating, over each small region of the scene, the product of the location-dependent, spectral reflectance curves against three particular frequency response curves. We refer to these three functionals as maps to red, green, and blue (RGB) space \cite{Westland2}.  Through an emulation of this map, a digital photograph represents a given scene under a given illumination as an $A\times B\times 3$ data array where the first two coordinates record the location in the scene and the last coordinate records the values of the red, green, and blue functionals over each pixel. By combining the three $A\times B$ color sheets, one can approximate a human's perception of the illuminated scene. 
 See Fig. \ref{fig:StillLifeSheets} for an illustration.

\begin{figure}
\centering
\subfloat[]{\label{fig:Sand1_3D}\includegraphics[width=.5\textwidth]{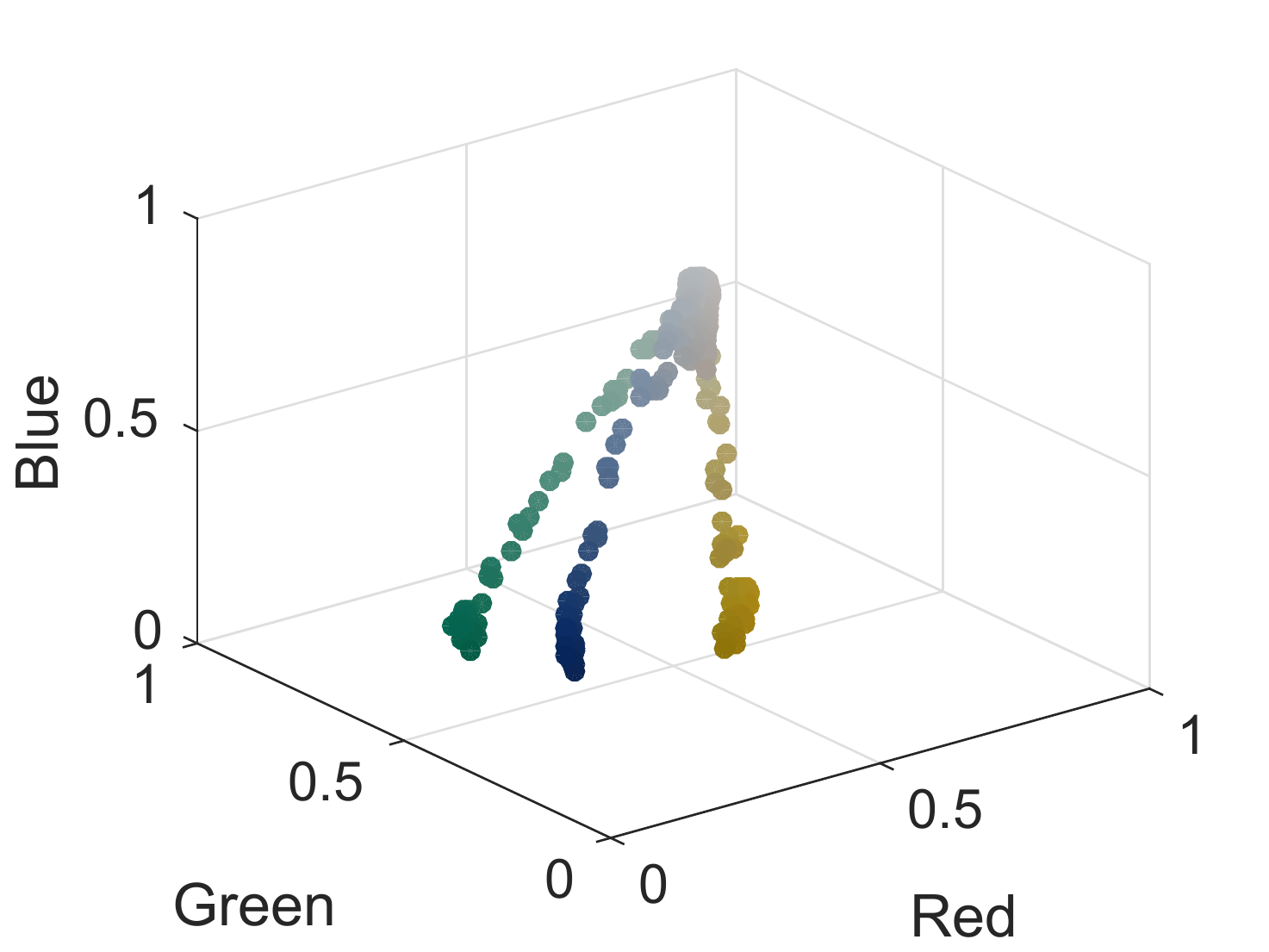}}
\subfloat[]{\label{fig:Sand2_3D}\includegraphics[width=.5\textwidth]{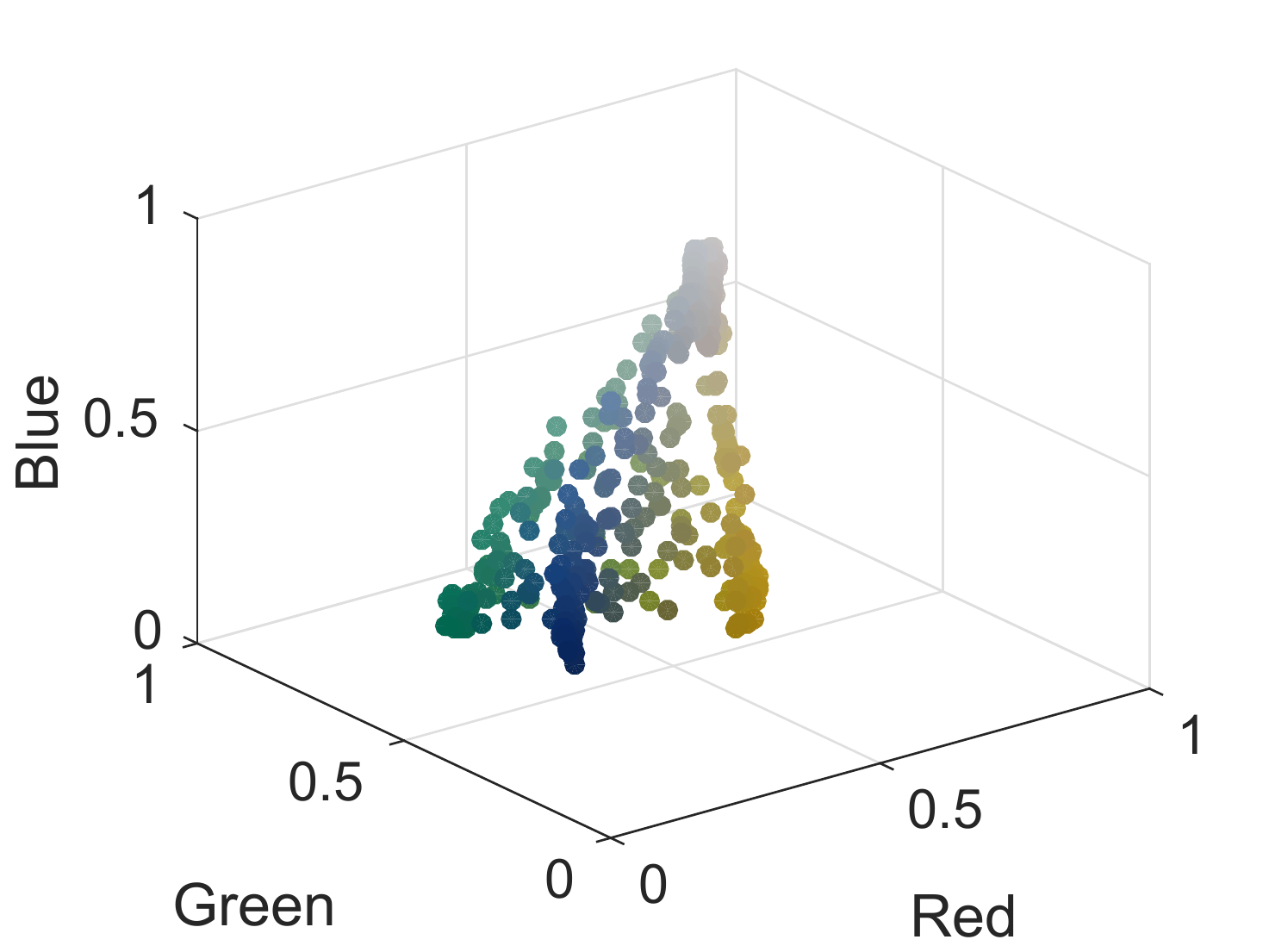}}\\
\subfloat[]{\label{fig:Sand3_3D}\includegraphics[width=.5\textwidth]{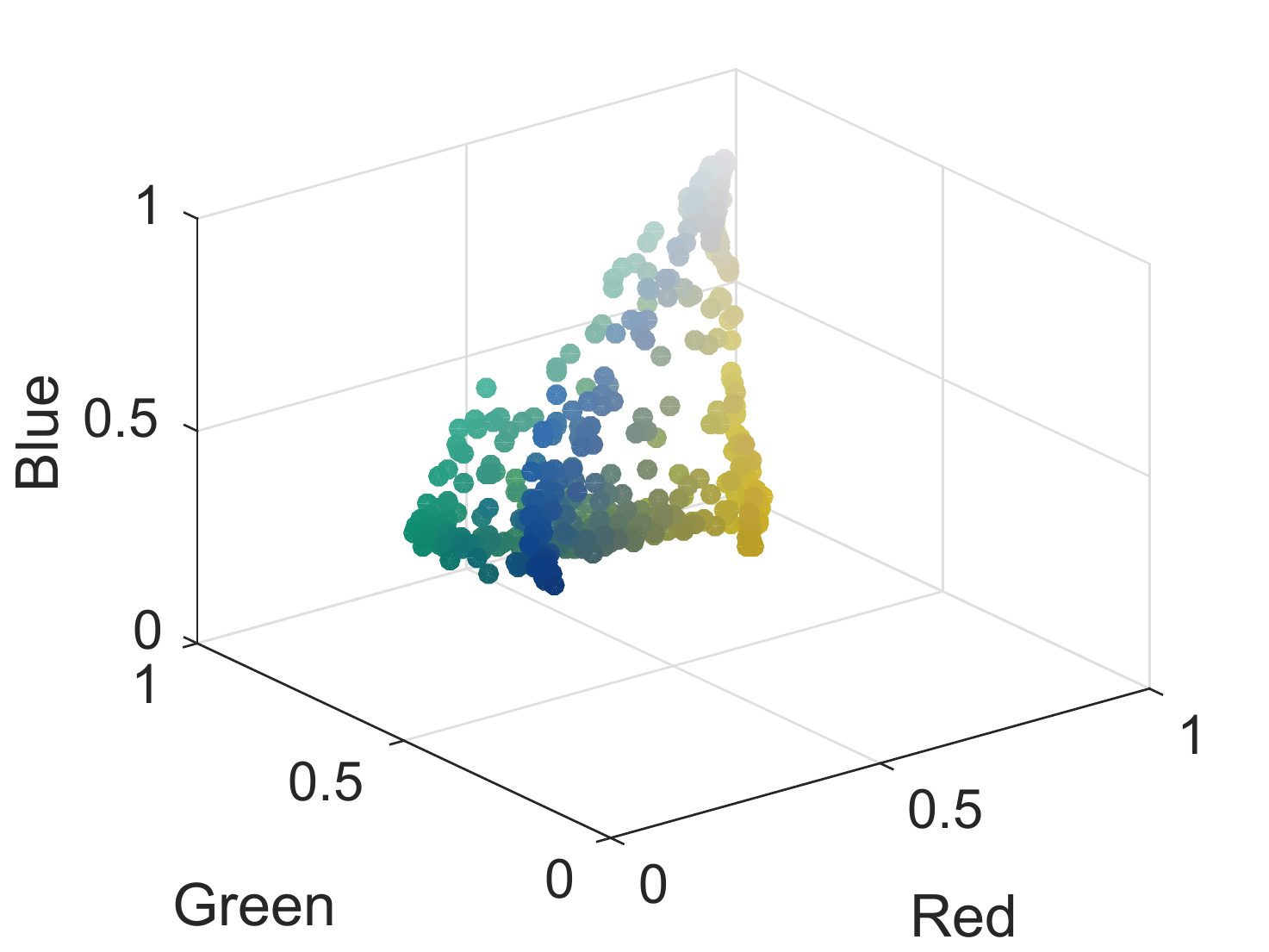}}
\subfloat[]{\label{fig:Sand4_3D}\includegraphics[width=.5\textwidth]{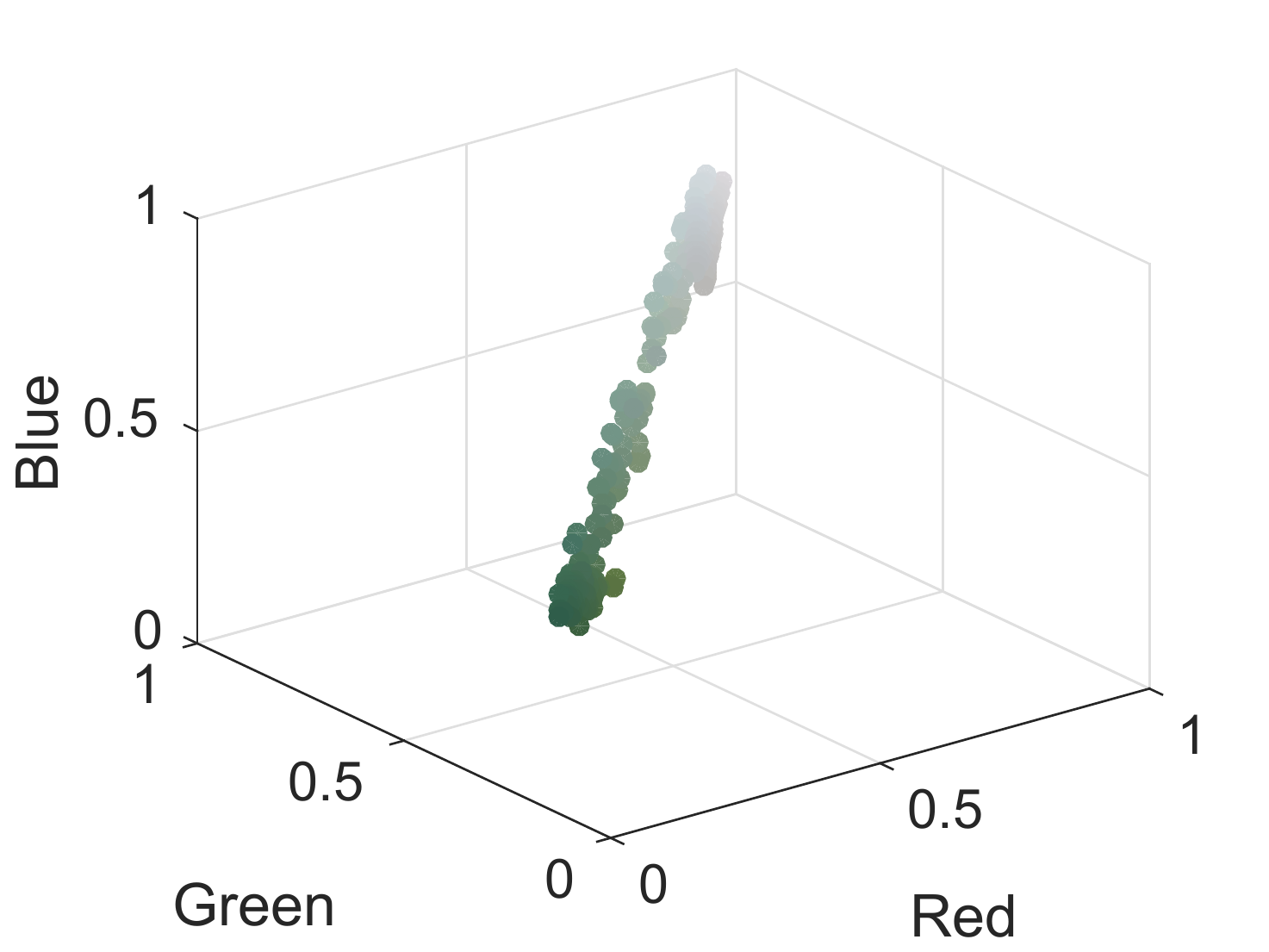}}
\caption{The RGB color data from each of the sand digital images with reduced resolution, colored according to each pixel value.}
\label{fig:SandRGB}
\end{figure}

The entries in the three $A\times B$ sheets correspond to the energy arriving near the red, green, or blue frequencies at each pixel.  The data set that we will consider is the family of $A B$ $3$-tuples present in a digital image.
For each of the four images in Fig. \ref{fig:Sand}, we reduce the resolution of the image and plot the 3-dimensional color data with each point colored according to the pixel value (see Fig. \ref{fig:SandRGB}).  The color data is normalized between $[0,1]$ for each color frequency where 0 indicates the absence of color and 1 indicates pure color. At first, one might suppose that the RGB color data from the image of unmixed sand would consist of four clusters of points in $\mathbb{R}^3$, one for each of the three colors of sand and another for the background.  However, we observe in Fig. \ref{fig:Sand1_3D} that there are three tendrils connecting points in $\mathbb R^3$ corresponding to each color of sand to the points in $\mathbb R^3$ corresponding to the background color. This is due to some pixels in the lower resolution image consisting not only of pure sand or pure background but a mixture of the two.  Fig. \ref{fig:Sand2_3D}--corresponding to the image of mixing two colors at a time, leaving some sand unmixed--results in a hollow tetrahedron with no base. In Fig. \ref{fig:Sand3_3D}, the addition of a region where all three colors of sand are mixed together, leaving some regions of unmixed and partially mixed sand, results in a solid tetrahedron.  Fig. \ref{fig:Sand4_3D} illustrates how a more complete mixing of the sand leads to a degeneration of the solid tetrahedron to a single tendril connecting the aggregate sand mixture to the background.

One might expect the convex hull of the point clouds in Figs. \ref{fig:Sand1_3D}-\ref{fig:Sand3_3D} to have four vertices corresponding to the four purest pixels within the data set.  However, a convex hull computation determines there to be roughly 40 vertices of the convex hull of the pixels in each image.  Consistent with intuition, the purest pixels in each of these images are extremal \emph{i.e.} among the extremal vertices of the convex hull corresponding to the point cloud of data.   The general identification of pure pixels has had compelling applications in archetypal analysis \cite{cutler} and endmember detection in hyperspectral imagery \cite{boardman}.  For instance, if a tract of land is imaged under a collection of $D$ spectral bands, then we can use the image to construct a data set in $\mathbb R^D$. The extremal points in this data set correspond to  locations captured within the image containing the potentially purest form of some mineral/substance. A desire to uncover such points and stratify all data points according to their proximity to the boundary motivated the development of the main algorithms in this paper.

\section{Background}
In this section, we provide background definitions and material for the main algorithm. While the following statements hold for any vector space $V$, we restrict our focus to a vector space over the reals $\mathbb{R}^D$. Given a set of vectors $C=\{\textbf{x}_1, \dots, \textbf{x}_p\} \subset \mathbb{R}^D$, an {\it affine combination} of the elements of $C$ is a vector $\textbf{x}=\sum_{j=1}^p w_j\textbf{x}_j$ with $\sum_{j=1}^p w_j=1$. A {\it convex combination} of the elements of $C$ is an affine combination with the additional constraint that $w_j\geq 0$ for all $j$. A subset $S\subset \mathbb{R}^D$ is {\it convex} if whenever $\textbf{x},\textbf{y}\in S$ and $t\in [0,1]$, we have $(1-t)\textbf{x}+t\textbf{y} \in S$.  In other words, the line segment joining any two points in $S$ also lies in $S$. As a consequence, the intersection of any collection of convex sets is convex. For $C$ a finite collection of points in $\mathbb{R}^D$, the {\it convex hull} of $C$ can be defined in several equivalent ways. It is the smallest convex set containing $C$, and it is the set of all convex combinations of points in $C$. We summarize these equivalent definitions below.
\begin{definition} 
Let $C\subset \mathbb{R}^D$ be a finite set of points. The \textit{convex hull}, $\mathcal H(C)$  of $C$, is:
\begin{itemize}
\item[i)] The smallest convex set containing $C$.
\item[ii)] $\mathcal{H}(C)=\{w_1 \textbf{x}_1+\cdots +w_{p} \textbf{x}_{p}\ |\ \textbf{x}_i \in C,\  w_i \geq 0,   w_1+\cdots+w_{p}=1\}.$
\end{itemize}
\end{definition}

An element $\textbf{v}\in C$ is a {\it vertex} of $\mathcal H(C)$\footnote{In 2-dimensions, $\mathcal H(C)$ is a convex polygon. In $D$-dimensions, $\mathcal H(C)$ is referred to as a {\it convex polytope}.} if and only if $\textbf{v}\notin \mathcal H(C\setminus {\textbf{v}})$. The following proposition is a characterization of a vertex of $\mathcal H(C)$ that lends itself to a better understanding of our approach. The proof is clear and is not included.
 \begin{proposition} \label{propch} If $C=\{\textbf{x}_1, \dots, \textbf{x}_p\}$, then $\textbf{x}_i$ is a vertex of $\mathcal H(C)$ if and only if there is exactly one solution to $\textbf{x}_i=\sum_{j=1}^pw_j\textbf{x}_j$ with $w_j \geq 0$ and  $\sum_{j=1}^p w_j=1$ (namely $w_j=0$ for $j\neq i$ and $w_i=1$).
 \end{proposition}

An element $\textbf{b} \in C $ is a {\it point on the boundary} of $C$ if and only if there exists an affine linear function $f(\textbf{y})=a_0+a_1y_1+\ldots+a_Dy_D$ with scalars $a_0,a_1,\ldots, a_D$  such that $f(\textbf{y})\geq 0$ for all $\textbf{y}\in C$ which also satisfies $f(\textbf{b})=0$. Note a vertex is also a point on the boundary. 
In the next section, we will be interested in identifying elements in $C$ that lie {\it on or near} the boundary of $\mathcal H(C)$. 

Consider a point cloud of data $C\subset \mathbb R^D$. We would like to represent each point $\textbf{x}_i\in C$, as accurately as possible, as an affine combination of its $K$ nearest neighbors as determined by some proximity relation. For instance, the $K$ nearest neighbors of point $\textbf{x}_i$ could be the $K$ points $\textbf{x}_j \neq \textbf{x}_i \in C$ which minimize $d(\textbf{x}_i,\textbf{x}_j)$ for some distance metric $d$. If possible, we would further like this representation to be a convex combination.  If the smallest affine linear space containing the $K$ nearest neighbors also contains $\textbf{x}_i$, then this affine representation will be exact.  

\begin{figure}
\centering
\includegraphics[width=\textwidth]{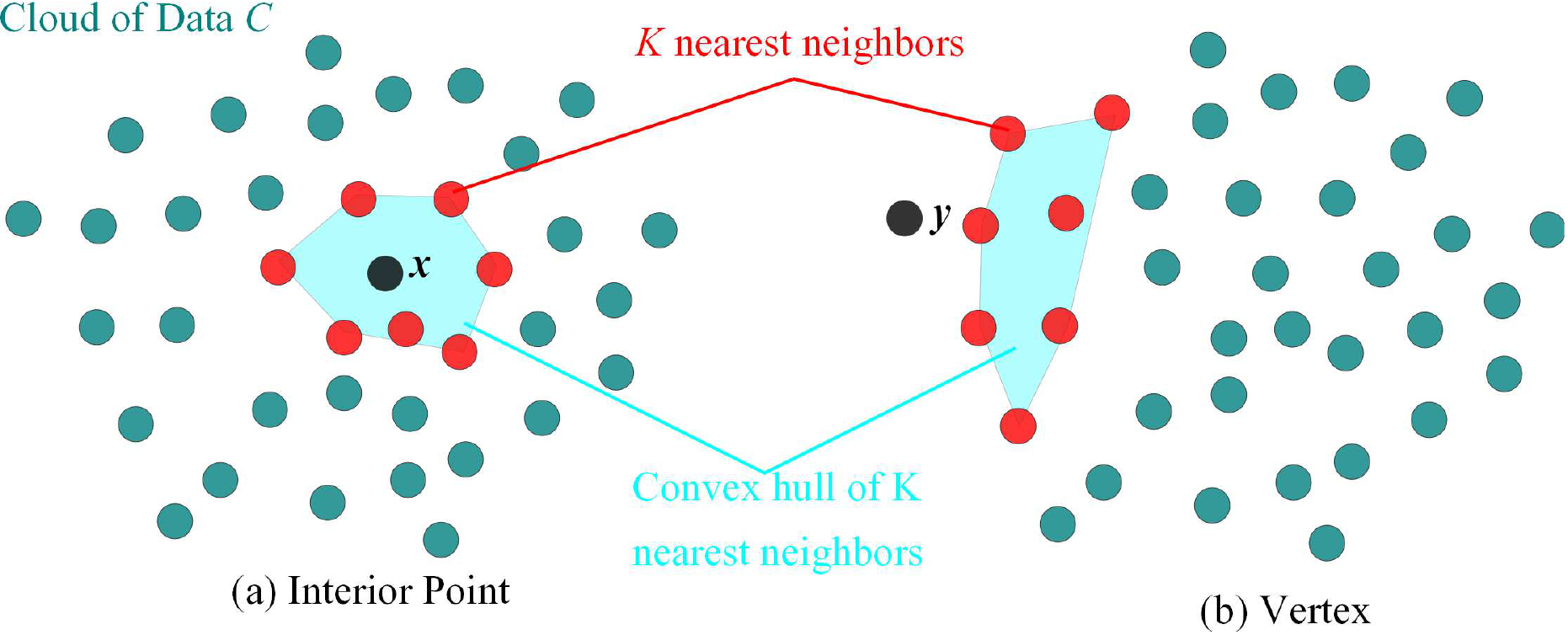}
\caption{Illustration of reconstructing an interior point versus a vertex by a set of $K=7$ nearest neighbors.}
\label{fig:ConvHullPoints}
\end{figure}

For instance, in the point cloud in Fig. \ref{fig:ConvHullPoints}, let $K=7$, and consider the black points $\textbf{x}, \textbf{y}$ lying in $\mathcal H(C)$. Both $\textbf{x}$ and $\textbf{y}$ lie in the affine span of their $7$ nearest neighbors. However,  $\textbf{x}$ falls within the convex hull of its  $7$ nearest neighbors. Thus,  we can reconstruct $\textbf{x}$ as an affine combination of its $7$ nearest neighbors such that all coefficients (or weights) are non-negative (see Fig. \ref{fig:ConvHullPoints}a).  The point $\textbf{y}$ does not fall within the convex hull of its nearest neighbors. Thus, while we can perfectly reconstruct $\textbf{y}$ as an affine combination of its neighbors, we cannot reconstruct it as a convex combination of its neighbors (see Fig. \ref{fig:ConvHullPoints}b). That is, when we reconstruct $\textbf{y}$ as an affine linear combination of its neighbors, we must allow at least one of the coefficients to be negative. In general, we have the following corollary to Proposition~\ref{propch}.

\begin{corollary}
Let $C=\{\textbf{x}_1, \dots, \textbf{x}_p\}$. If $\textbf{x}_i$ is in the affine span of $C\setminus \{\textbf{x}_i\}$ then $\textbf{x}_i$ is a vertex of $\mathcal H(C)$  if and only if any exact representation of $\textbf{x}_i$ as an affine combination of the elements of $C\setminus \{\textbf{x}_i\}$ requires at least one negative coefficient.
\end{corollary} 

With this result in mind, we proceed to an optimization problem to determine both the vertices of $\mathcal H(C)$ and the set of points on or near the boundary of $\mathcal H(C)$.

\section{The Optimization Problem}
In this section, we discuss an optimization problem for determining  which elements of a point cloud $C=\{\textbf{x}_1, \dots, \textbf{x}_p\}\subset \mathbb{R}^D$ are close to the boundary of $\mathcal H(C)$. Let $N_i$ denote the indices of the set of {\it neighbors} of an element $\textbf{x}_i\in C$. For instance, $N_i$ could be the indices of the set of $K$ nearest neighbors to $\textbf{x}_i$ (\textit{i.e.} the $K$ indices corresponding to the $\textbf{x}_j \neq \textbf{x}_i$ with the smallest distance $d(\textbf{x}_i,\textbf{x}_j)$ for some metric $d$) or the indices of the set of elements in $C$ that are within some prescribed distance to $\textbf{x}_i$. Let $\textbf{w}_i=(w_{i1}, \dots, w_{ip})$ denote a {\it weight vector} associated to $\textbf{x}_i$ and let $\gamma, \lambda$ be a pair of non-negative real parameters. For each $\textbf{x}_i\in C$, consider the following optimization problem:
\begin{framed}\label{main}
\begin{equation}
\begin{aligned}
& \underset{\textbf{w}_i}{\text{minimize}} & & \gamma \|\textbf{w}_i\|_2^2 + \lambda \|\textbf{w}_i\|_1+ \| \textbf{x}_i-\displaystyle\sum_{j\in{N_i}}w_{ij}\textbf{x}_j \|_2^2  \\
& \text{subject to} & & \displaystyle \sum_{j\in{N_i}}w_{ij} =1. 
\end{aligned}
\label{eq:endmemberOp}
\end{equation}
\end{framed}
\noindent

Without the affine constraint $\sum_{j\in{N_i}}w_{ij} =1$, this optimization problem corresponds to elastic net regularization, which is known for its sparsity inducing properties \cite{Zou}.  In the setting above, the $\ell_1$ term induces convexity rather than sparsity. The benefits of the $1$-norm have been widely recognized in signal and image processing, see \emph{e.g.} \cite{candes2006robust,goldstein2009split,boyd2011distributed}.  We note that the third term in the objective function has been proposed as one step in the dimensionality reduction algorithm {\it local linear embedding} \cite{roweis2000nonlinear} which is not concerned with the convex hull stratification problem being addressed here. In general, convex  optimization problems are attractive because of the guaranteed existence of a global minimum and a spectrum of available numerical solvers \cite{Boyd}.  
 
The solution to optimization problem \ref{eq:endmemberOp} consists of a representation of $\textbf{x}_i$ as a $K$-dimensional 
weight vector $\textbf{w}_i$, where the possible non-zero weights of $\textbf{w}_i$ are supported on $N_i$.  We now describe how this weight vector encodes geometric information about the proximity of each data point to the boundary of $\mathcal H(C)$ by analyzing the role of each term in the objective function.


\subsection{Data representation}
As a starting point, consider the role of the third term of the optimization problem:
\begin{framed}
\begin{equation}
\begin{aligned}
& \underset{\textbf{w}_i}{\text{minimize}} & & \| \textbf{x}_i-\displaystyle\sum_{j\in{N_i}}w_{ij}\textbf{x}_j \|_2^2  \\
& \text{subject to} & & \displaystyle \sum_{j\in{N_i}}w_{ij} =1. 
\end{aligned}
\label{eq:Residual}
\end{equation}
\end{framed}
\noindent
The goal of this term is to represent $\textbf{x}_i$ as an affine combination of the elements in $N_i$ with as little error as possible, with respect to the standard Euclidean norm.
If $|N_i|>D$ (where $D$ is the dimension of the ambient vector space where the data resides), then there will be infinitely many weight vectors minimizing this term. This non-uniqueness allows further terms in the objective function to uncover solutions with additional properties.  

\subsection{Positivity}
Now consider the role of the second term in the optimization problem:
\begin{framed}
\begin{equation}
\begin{aligned}
& \underset{\textbf{w}_i}{\text{minimize}} & & \|\textbf{w}_i\|_1 \\
& \text{subject to} & & \displaystyle \sum_{j\in{N_i}}w_{ij} =1. 
\end{aligned}
\label{eq:1norm}
\end{equation}
\end{framed}
\noindent
Since $1=\sum_{j\in N_i}w_{ij}\leq  \sum_{j\in N_i}|w_{ij}|=\|\textbf{w}_i\|_1$, a minimal solution strives to have $\|\textbf{w}_i\|_1=1$. In such a solution, all of the entries of the weight vector will be non-negative.   
Thus, this term of the optimization problem seeks to
restrict the representation of $\textbf{x}_i$ to be a
convex combination of points in $C$.  In other words, this positivity term
induces convexity. 

\subsection{Uniformity}

Now consider the role of the first term in the optimization problem:

\begin{framed}
\begin{equation}
\begin{aligned}
& \underset{\textbf{w}_i}{\text{minimize}} & & \|\textbf{w}_i\|_2^2 \\
& \text{subject to} & & \displaystyle \sum_{j\in{N_i}}w_{ij} =1. 
\end{aligned}
\label{eq:2Norm}
\end{equation}
\end{framed}
\noindent
Using Lagrange multipliers, one can show that the solution strives to satisfy
$w_{ij} = 1/|N_i|$
 for all $j\in N_i$.  Hence, an optimal solution seeks uniformity in the weights. This term also acts to regularize the problem so that it has a unique solution.

\subsection{Putting it all together}
We now consider the role of $\gamma$ and $\lambda$ in the optimization problem.   We seek to represent
a point as an affine linear combination of its neighbors,  we would like that all the weights are non-negative, and we
favor solutions with a uniform distribution of weights.  

If we choose the parameters such that $\gamma<\lambda<1$\footnote{This choice of parameters assumes that the input data $\textbf{x}_i$ is not too small in magnitude. A discussion about the scale of the data and how it affects parameters follows momentarily.}, then primary emphasis is placed on reconstructing each data point as best as possible (the third term), secondary emphasis is placed on finding weights reflecting a convex combination if possible (the second term), and tertiary emphasis is placed on uniformity in the coefficients of the weight vector (the first term). The uniformity component also serves to regularize the optimization problem. That is, it serves to determine a unique solution if the first two steps lead to many possible solutions.  Therefore, each term in the objective function enforces weights favoring (from right to left) representation, convexity, and uniformity subject to the constraint of being an affine combination.
\begin{equation}
\begin{aligned}
& \underset{\textbf{w}_i}{\text{minimize}} & & \overbrace{\gamma \|\textbf{w}_i\|_2^2}^\text{Uniformity} + \overbrace{\lambda \|\textbf{w}_i\|_1^{\phantom{i}}}^\text{Convexity}+ \overbrace{\| \textbf{x}_i-\displaystyle\sum_{j\in{N_i}}w_{ij}\textbf{x}_j \|_2^2}^\text{Representation} \\
& \text{subject to} & & \underbrace{\displaystyle \sum_{j\in{N_i}}w_{ij} = 1.}_\textrm{Affine Combination}  
\end{aligned}
\label{eq:endmemberOpExplain}
\end{equation}

If $\gamma\ll \lambda\ll 1$, then representability is essentially enforced as a first step. If there is a unique solution to the affine representation problem then the additional terms in the optimization condition, involving $\lambda \|\textbf{w}_i\|_1$ and $\gamma \|\textbf{w}_i\|_2^2$, are effectively ignored. If there are multiple solutions to the affine reconstruction problem, then from among these solutions, the weight vectors with the smallest $\ell_1$ norm are selected. If this yields a unique answer, then the third term, involving $\gamma \|\textbf{w}_i\|_2^2$, is effectively ignored. If there are multiple solutions to both the affine representation problem and to the minimized $\ell_1$ norm problem, then the solution with the smallest $\ell_2$ norm is selected. If $\textbf{x}_i$ is in the affine span of $C\setminus \{\textbf{x}_i\}$ and if $N_i$ includes the index of every element in $C$ other than $i$, then solving this optimization problem will result in $\textbf{w}_i$ having a negative component if and only if $\textbf{x}_i$ is a vertex of the convex hull. The $\ell_2$ norm of $\textbf{w}_i$ measures the bias of $\textbf{x}_i$ relative to the elements indexed by $N_i$, which we will see illustrated in the examples below.

Notice that if a set of data is scaled by constant $\alpha$, then the third term in optimization problem \ref{eq:endmemberOpExplain} will be scaled by $\alpha^2$. Therefore, given two sets of data represented as $D \times p$ matrices $X$ and $Y$ where $Y=\alpha X$ there are two analogous approaches to be used to optimize for the decision variables $\textbf{w}_i$ corresponding to data set $Y$: (1) $Y$ can be scaled to lie within the same range as $X$, and then the same parameters $\lambda$ and $\gamma$ used to optimize for $X$ should be used for the rescaled data, or (2) the parameters $\lambda$ and $\gamma$ used to optimize for $X$ can be scaled by $\alpha^2$ to optimize for the data $Y$. For the examples in this paper, we have chosen the first approach where our data has been scaled between 0 and 1. We observe that despite different numbers of points and different ambient dimensions of the data, comparable choices of parameters $\lambda$ and $\gamma$ have similar effects on the optimization problem.

\section{The Convex Hull Stratification Algorithm} \label{sec:algorithm}
In this section, we reformulate optimization problem (\ref{eq:endmemberOp}) as a quadratic program of the form
\begin{equation}
\begin{aligned}
& \underset{\textbf{w}}{\text{minimize}} & & \frac{1}{2}\textbf{w}^T Q \textbf{w} + \textbf{c}^T\textbf{w} \\
& \text{subject to} & & A\textbf{w} =  \textbf{b}, \ \  \textbf{w}      \geq  0.\\
\end{aligned}
\label{eq:endmemberOpQuad}
\end{equation}
We introduce the nonnegative variables $w_j^+$ and $w_j^-$ such that $w_j=w_j^+-w_j^-$ and $|w_j|=w_j^++w_j^-$ (as suggested in \cite{bertsimas}) to rewrite the non-differentiable portion of the objective function. We then expand and group the quadratic and linear terms in the vector $\textbf{w}$.  Note that the constant term is neglected as it does not affect the optimal solution to this problem.  The quadratic program is in $N=2K$ variables with  $A$ a  $1 \times 2K$ matrix.  This is a convex optimization problem for which there are many solvers, \emph{e.g.} \cite{Boyd,goldstein2009split,boyd2011distributed}. The selection of the convex solver is somewhat arbitrary given the guaranteed existence of a global minimum. In the examples presented in this paper, we choose to optimize using the Primal Dual Interior Point method as described in \cite{bertsimas,Boyd,vanderbei}.  We solve problem \ref{eq:endmemberOp} by reducing it to a sequence of linear, equality constrained problems and then applying Newton's method. For further details, see \cite{Ziegelmeier}.  

\begin{algorithm}
\KwData{points $\textbf{x}_i \in C$, a cloud of data}
\KwResult{weight vector $\textbf{w}_i$ associated to each point}
initialize parameters $\gamma$ and $\lambda$\;
\For{ each $\textbf{x}_i\in C$}{
determine indices $N_i$ of the $K$ nearest neighbors to $\textbf{x}_i$\;
solve 
\[
\begin{aligned}
&\underset{\textbf{w}_i}{\textrm{minimize}} & \gamma \|\textbf{w}_i\|_2^2 + \lambda \|\textbf{w}_i\|_1+ \| \textbf{x}_i-\displaystyle\sum_{j\in{N_i}}w_{ij}\textbf{x}_j \|_2^2 \\
&\textrm{subject to} & \quad \displaystyle\sum_{j\in{N_i}}w_{ij} =1;  \\
\end{aligned}\;
\]
determine if $\textbf{w}_i$ has a negative entry\;
compute $||\textbf{w}_i||_2$\;
}
\caption{Convex Hull Stratification Algorithm to determine proximity of data in relation to the convex hull. The weight vector $\textbf{w}_i$ encodes geometric information that leads to a stratification of the data, illustrated in the examples below.}
\label{alg}
\end{algorithm}

The algorithm is summarized in Algorithm \ref{alg} and illustrated in the context of examples in Section \ref{sec:Examples}. To estimate the computational complexity, note that determining the nearest neighbors scales like $\mathcal{O}(Dp^2)$ where $D$ is the ambient dimension of the input data, and $p$ is the number of data points.  Formulating the quadratic program as a linear system of equalities is on the order of $\mathcal{O}(K^2Dp)$.  Solving the full $(2N+M) \times (2N+M)=(4K+1) \times (4K+1)$ system of equations scales like $\mathcal{O}(p(4K+1)^3h)$, where $h$ is the average number of iterations required to solve each problem. 
 Note that each step can be solved independently for each point $\textbf{x}_i$, thus the algorithm is trivially parallelizable. As previously mentioned, if $K=p-1$ (that is, the neighbor set $N_i$ of point $\textbf{x}_i$ contains all other points in the data set), then the corresponding weight vector $\textbf{w}_i$ will have a negative component if and only if $\textbf{x}_i$ is a vertex of the convex hull. In practice, we often choose $K$ to be much smaller than $p-1$ to speed up computations.

\section{Examples} \label{sec:Examples}
In this section, we illustrate the use of the algorithm with several different data sets ranging from illustrative toy problems to the analysis of hyperspectral data in $\mathbb{R}^{20}$ and real image data from the sand experiment. We show how selecting the parameters $\gamma$ and $\lambda$ can provide two different approaches for detecting proximity to the convex hull.

\subsection{Polygon in the plane} \label{sec:Polygon}

\begin{figure*}
  \centering
  \subfloat[$\lambda=10^{-5}, \gamma=10^{-6}$]{\label{fig:l5g6} \includegraphics[width=.5\textwidth]{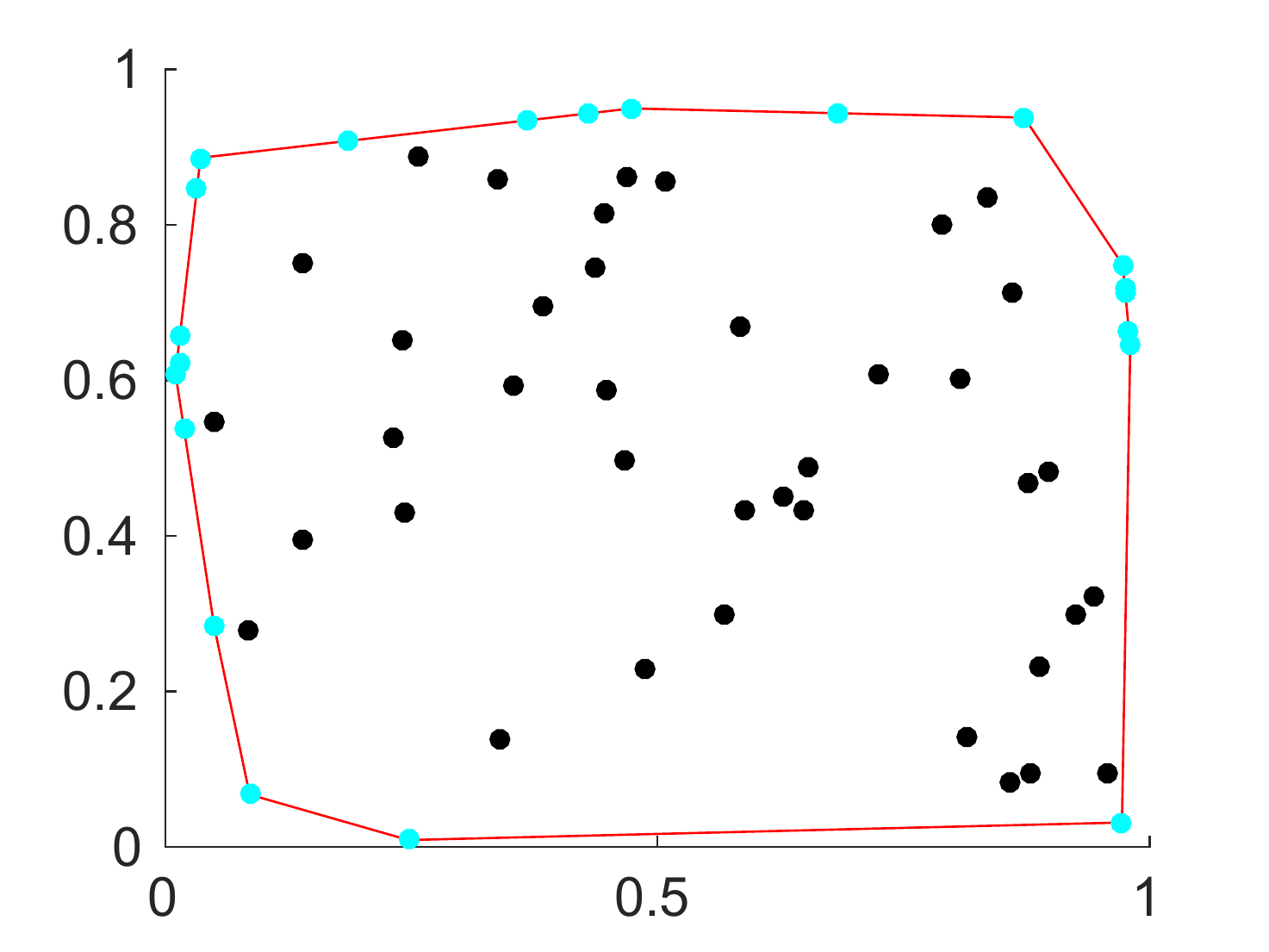}}
  \subfloat[$\lambda=10^{-5}, \gamma=10^{-3}$]{\label{fig:l5g3} \includegraphics[width=.5\textwidth]{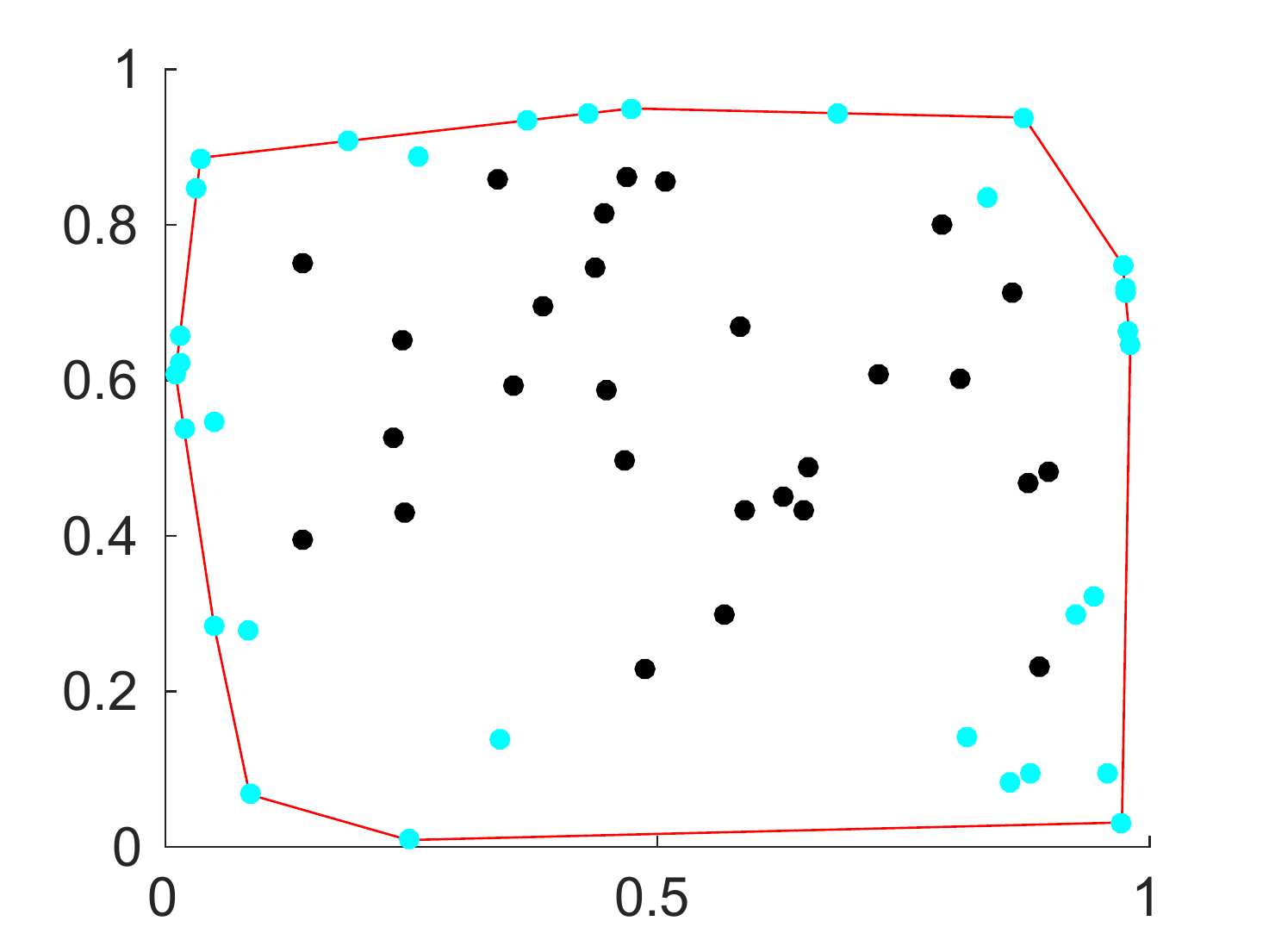}}\\
  \subfloat[$\lambda=10^{-3}, \gamma=10^{-6}$]{\label{fig:l3g6} \includegraphics[width=.5\textwidth]{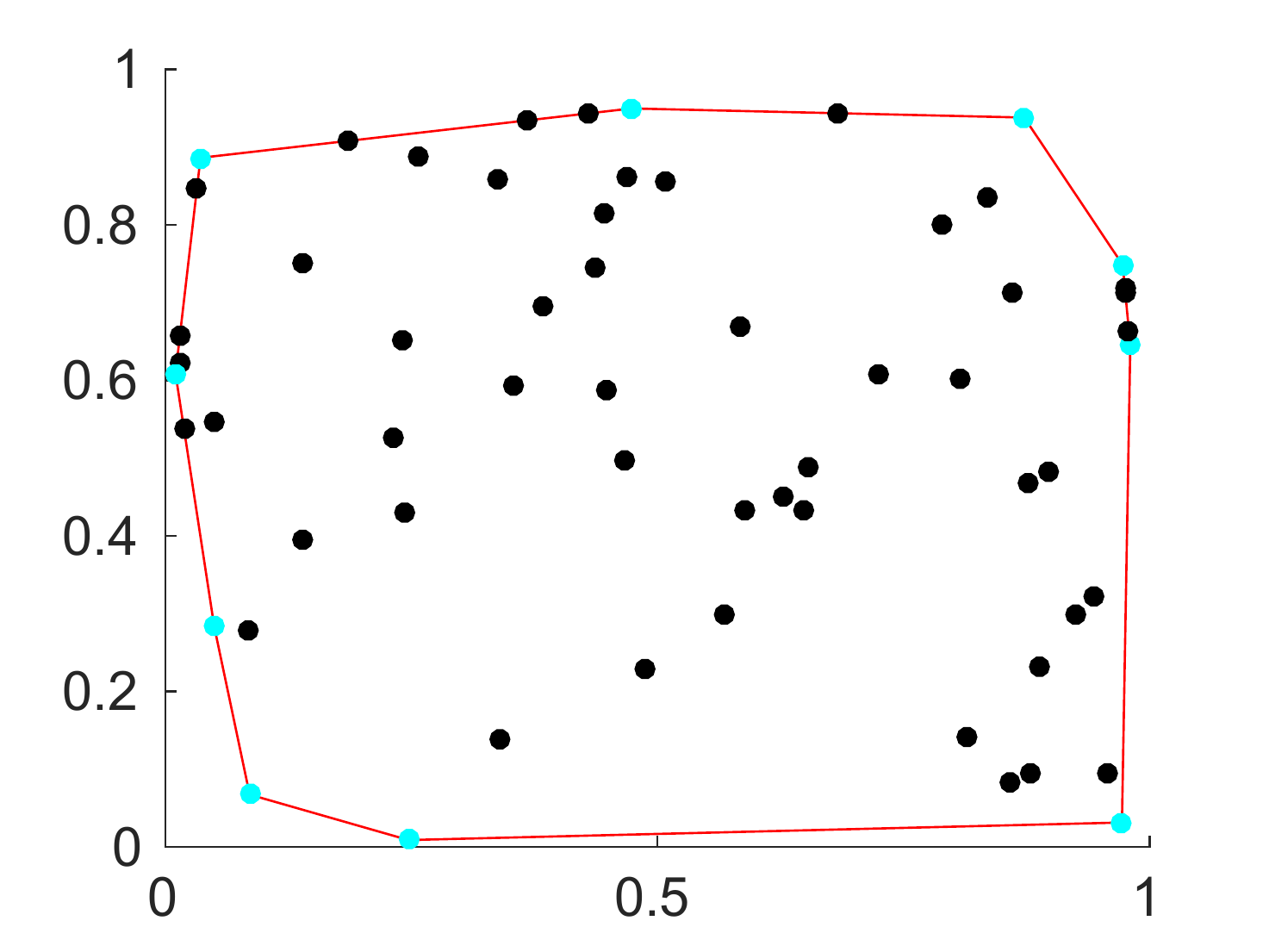}}
  \subfloat[$\ell_2$ Norms of Weights]{\label{fig:weightspoly} \includegraphics[width=.5\textwidth]{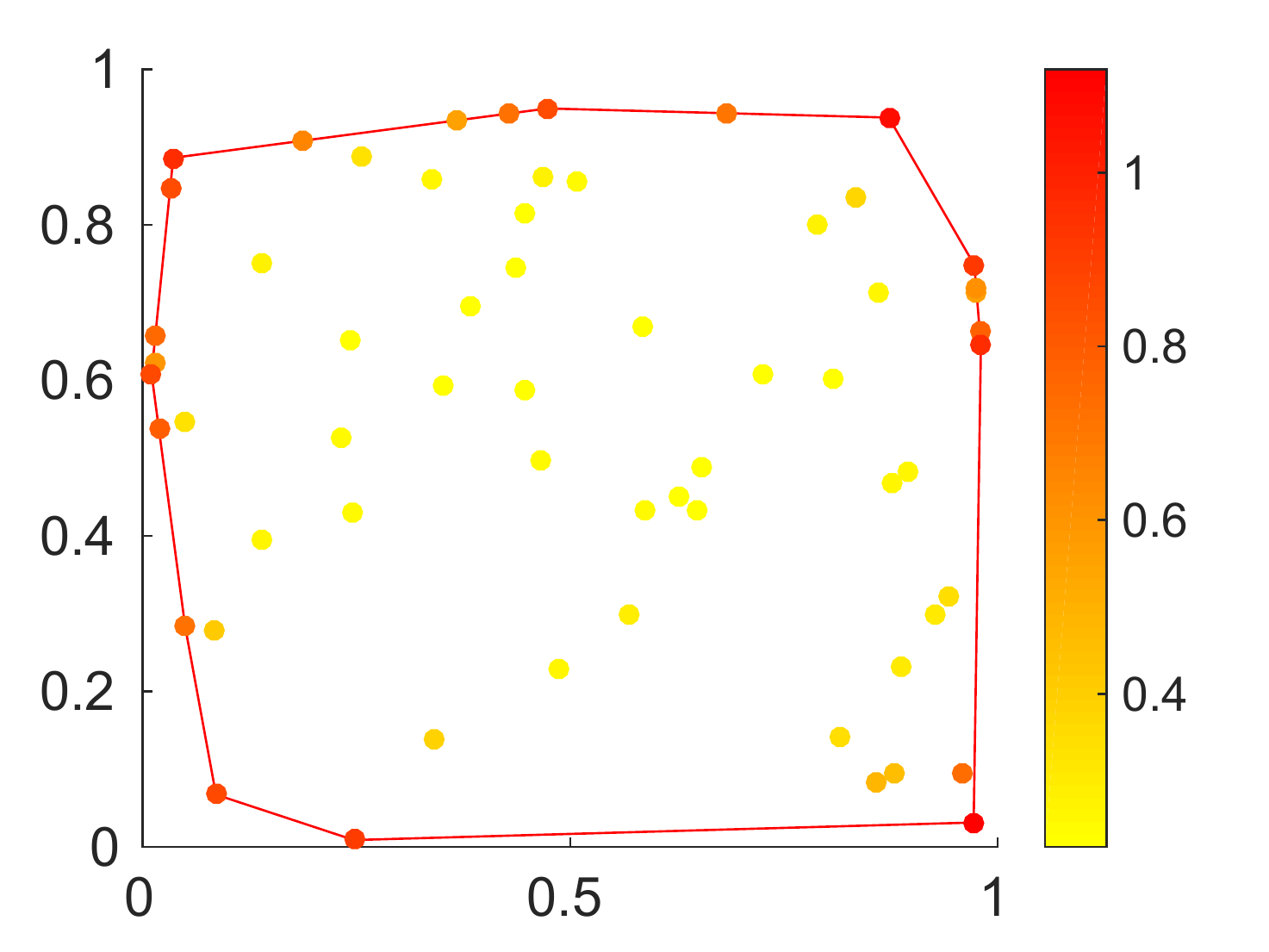}}
  \caption{(a)-(c) Implementation of algorithm with various parameter choices on points in the plane.  Points represented in cyan have a negative weight. (d) Data points colored according to the magnitude of the $\ell_2$ norm of the weight vector (using parameters $\lambda=10^{-3}, \gamma=10^{-6}$), where yellow corresponds to the smallest values and red corresponds to the largest values.}
  \label{fig:EucConvexHull}
\end{figure*}

In the first example, we generate 50 uniformly distributed random points inside the unit square and place an additional 10 points along the boundary edges of the convex hull for a total of 60 points.  We select $K=20$ nearest neighbors using a metric of Euclidean distance for proximity (this measure is adopted throughout all experiments in this paper).  Fig. \ref{fig:EucConvexHull}, shows three examples with $\lambda,\gamma \ll 1$. Varying the relative values of $\lambda$ and $\gamma$ illustrates how the optimization trades a desire for convexity with a desire for uniformity.  Points whose weight vectors have a negative term are illustrated as cyan and points with only non-negative weights as black.  With the choice of $\lambda=10^{-5}$ and $\gamma=10^{-6}$, we see in Fig. \ref{fig:l5g6} that all of the boundary points have negative weights yielding a thin cyan boundary shell.  If a thicker boundary is desired, then a choice of $\gamma$ and $\lambda$ as seen in Fig. \ref{fig:l5g3} would be appropriate.  As more emphasis is placed on uniformity than on convexity, more weights become negative.  In Fig. \ref{fig:l3g6}, a strong emphasis is placed on convexity and only the vertices of the convex hull are represented using negative weights. The thickness of the boundary shell uncovered by the algorithm is controlled by varying the relative values for $\lambda$ and $\gamma$. One can thus stratify points near the boundary of the convex hull by varying the relative values of $\lambda$ and $\gamma$ and recording the emergence of weight vectors with negative coefficients.

\subsection*{Weights encode geometric information}
The structure of the optimization problem also leads to an alternate stratification in the case where $\gamma \ll \lambda \ll 1$. This stratification results from the encoding of geometric information in the coefficients of the weight vector. One can then infer a considerable amount of information about the location of a data point relative to its neighbors based on the $\ell_2$-norm of its associated weight vector.
In particular, when a point is close to the boundary, it tends to lie near the edge of the convex hull of its nearest neighbors. This boundary effect  biases the relationship of $\textbf{x}_i$ to its nearest neighbors which leads to less uniformity in the weight coefficients. This bias can be measured by $\|\textbf{w}_i\|_2$. When a point is not near the boundary, it tends to lie near the average of its nearest neighbors and is well expressed by a uniform weight vector. In particular, the less uniform the weights, the larger the $\ell_2$-norm of the weight vector and the closer the point is to the boundary. In other words:  {\it The $\ell_2$-norm of the weight vector is a measure of the proximity of the associated 
point to the boundary.} This yields a method for coarsely ordering points in the data set based on proximity to the boundary of the convex hull.

In Fig. \ref{fig:weightspoly}, we color each point in the data set according to the magnitude of $\|\textbf{w}\|_2$ of its associated weight vector (as found in the optimization with parameters $\lambda=10^{-3}$ and $\gamma=10^{-6}$). Yellow corresponds to the smallest values and red corresponds to the largest values.  Thus, we see that the vertices have the largest Euclidean norm, followed by points along the boundary of the convex hull, followed by points near the boundary, then finally a general mass of yellow interior points.  In practice, the Euclidean norms of the weights are robust with respect to the optimization parameters. This norm allows for a useful stratification of the data. Therefore, we call our method the Convex Hull Stratification Algorithm (CHSA).

\subsection{Distributions in the Plane}

\begin{figure*}
  \centering
   \includegraphics[width=0.49\textwidth]{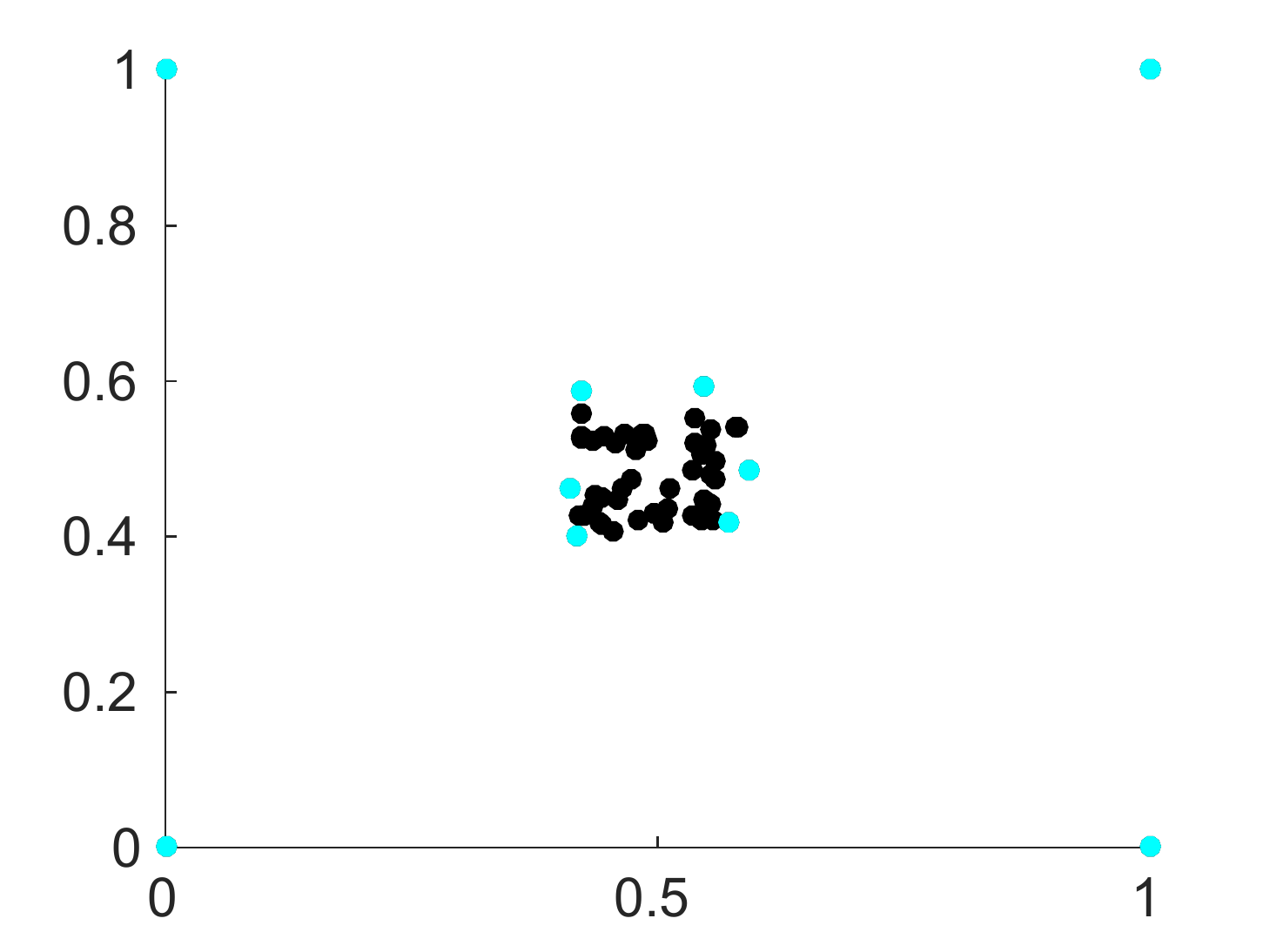}
   \includegraphics[width=0.49\textwidth]{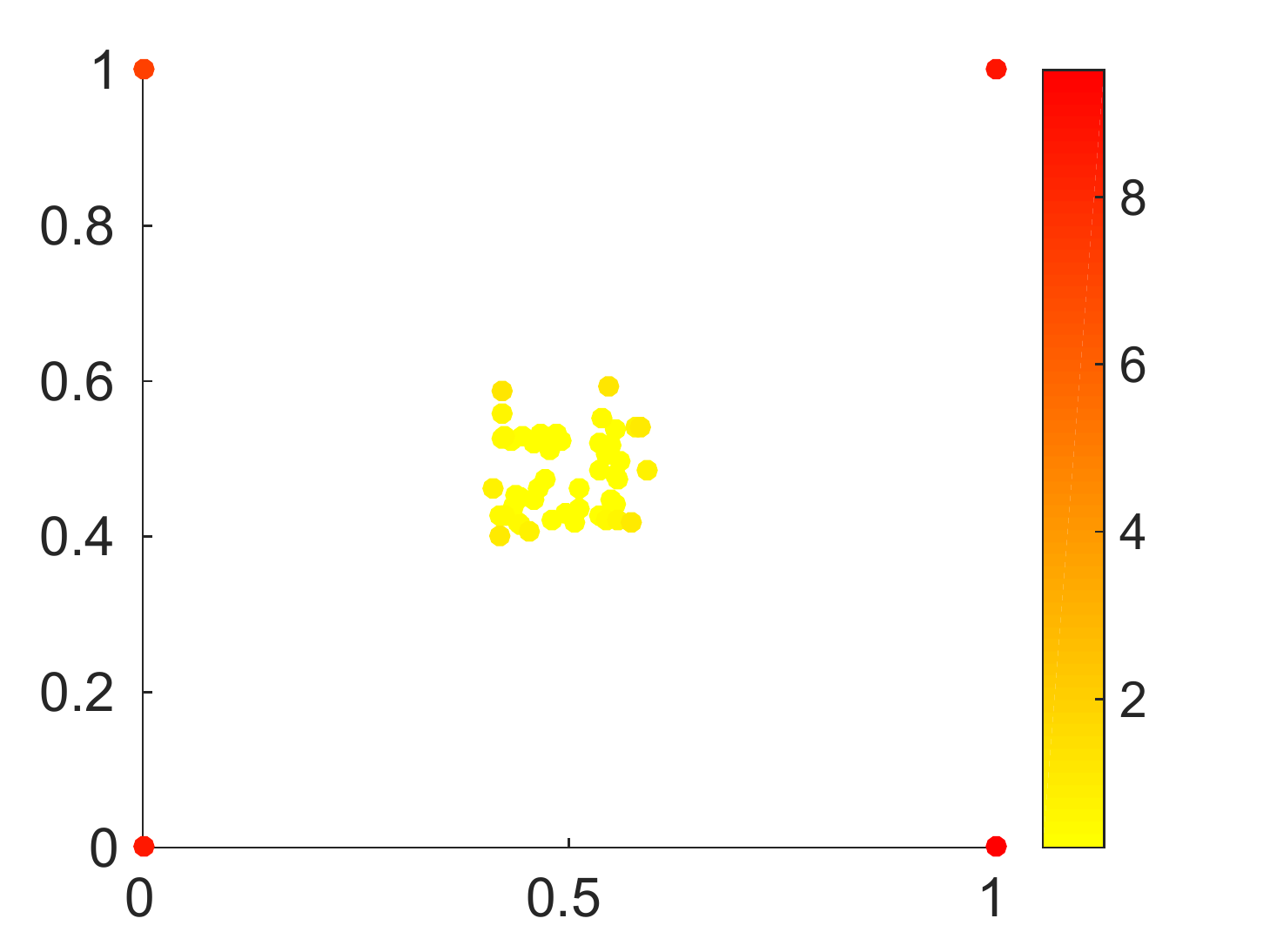}
	\caption{Distribution of points in the plane analyzed with CHSA using parameters $K=20$ nearest neighbors, $\lambda=10^{-3}$, and $\gamma=10^{-6}$. The left plot displays vertices of the convex hull (in cyan) with negative weights, and the right plot displays the points colored according to the norm of their weight vector.}
	\label{fig:Example1K20}
\end{figure*}

\begin{figure*}
  \centering
   \includegraphics[width=0.49\textwidth]{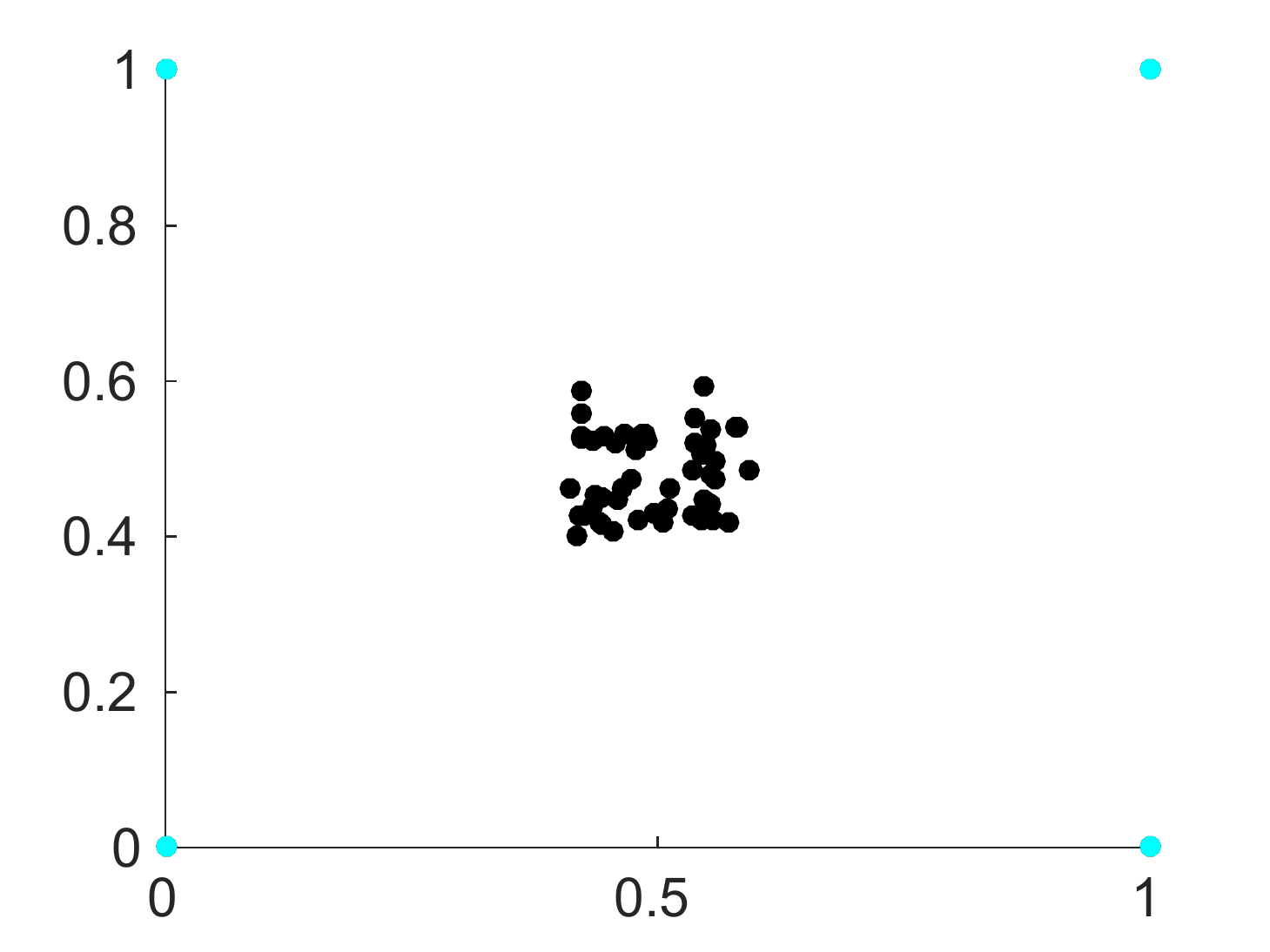}
    \includegraphics[width=0.49\textwidth]{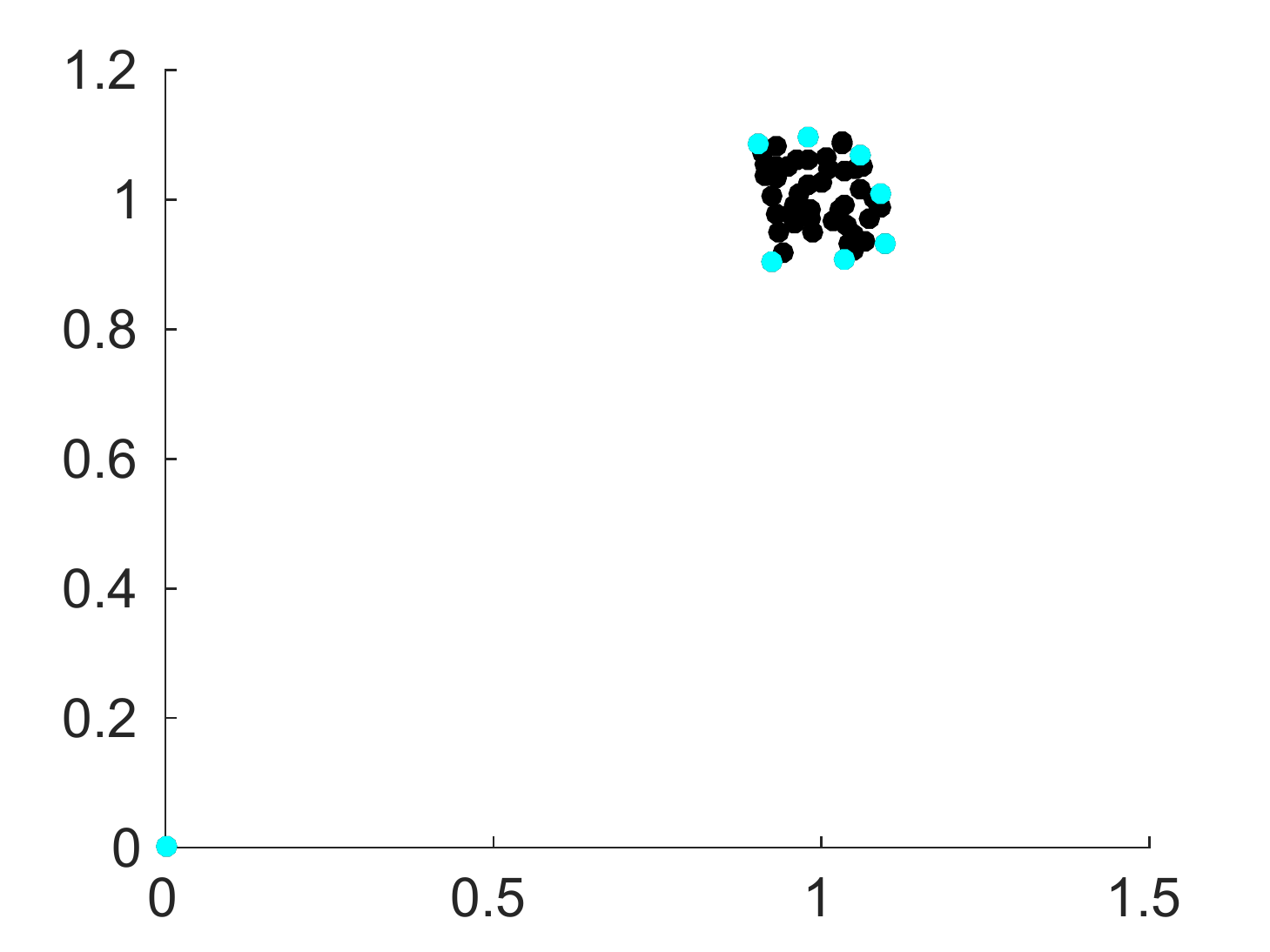}
	\caption{Distribution of points in the plane analyzed with CHSA using parameters $K=p-1$ nearest neighbors, $\lambda=10^{-3}$, and $\gamma=10^{-6}$. The left plot displays vertices of the convex hull (in cyan) with negative weights, and the right plot displays the points colored according to the norm of their weight vector.}
	\label{fig:Example1K53}
\end{figure*}

In this section, we discuss the algorithm's performance on examples in the plane with a variety of distributions. The first example is a set of data with one point in each corner of the unit square and 50 other uniformly distributed points clustered at the center $(0.5,0.5)$. For this and the following simulations, we display a figure with two subplots which reveal (on the left) the candidate vertices determined by our algorithm colored in cyan and (on the right) the points colored according to the magnitude of the norm of their weight vector. In Fig. \ref{fig:Example1K20}, we have selected the same parameters as in Figs. \ref{fig:l3g6} and \ref{fig:weightspoly}, $K=20$ nearest neighbors, $\lambda=10^{-3}$, and $\gamma=10^{-6}$. We observe that our algorithm reveals the four corners of the unit square as well as candidate vertices of the convex hull of the interior cluster. We also observe that there is a stark difference between the norms of the weight vectors associated to the points on the four corners as compared to points in the interior cluster. Our algorithm reveals useful information beyond that of a standard convex hull algorithm for two reasons: (1) it is not only able to uncover the four corners as vertices of the convex hull but also candidates within the interior cluster (2) the norm of the weight vector leads to a stratification that reveals the four corners as clear outliers as well as distinguishes the vertices of the interior cluster. If only the true vertices of the convex hull are desired, changing the choice of nearest neighbors to $K=p-1$ (as suggested at the end of Section \ref{sec:algorithm}) will reveal this behavior. See Fig. \ref{fig:Example1K53} where only the 4 corners of the unit square have negative weights using CHSA.

\begin{figure*}
  \centering
  \includegraphics[width=0.49\textwidth]{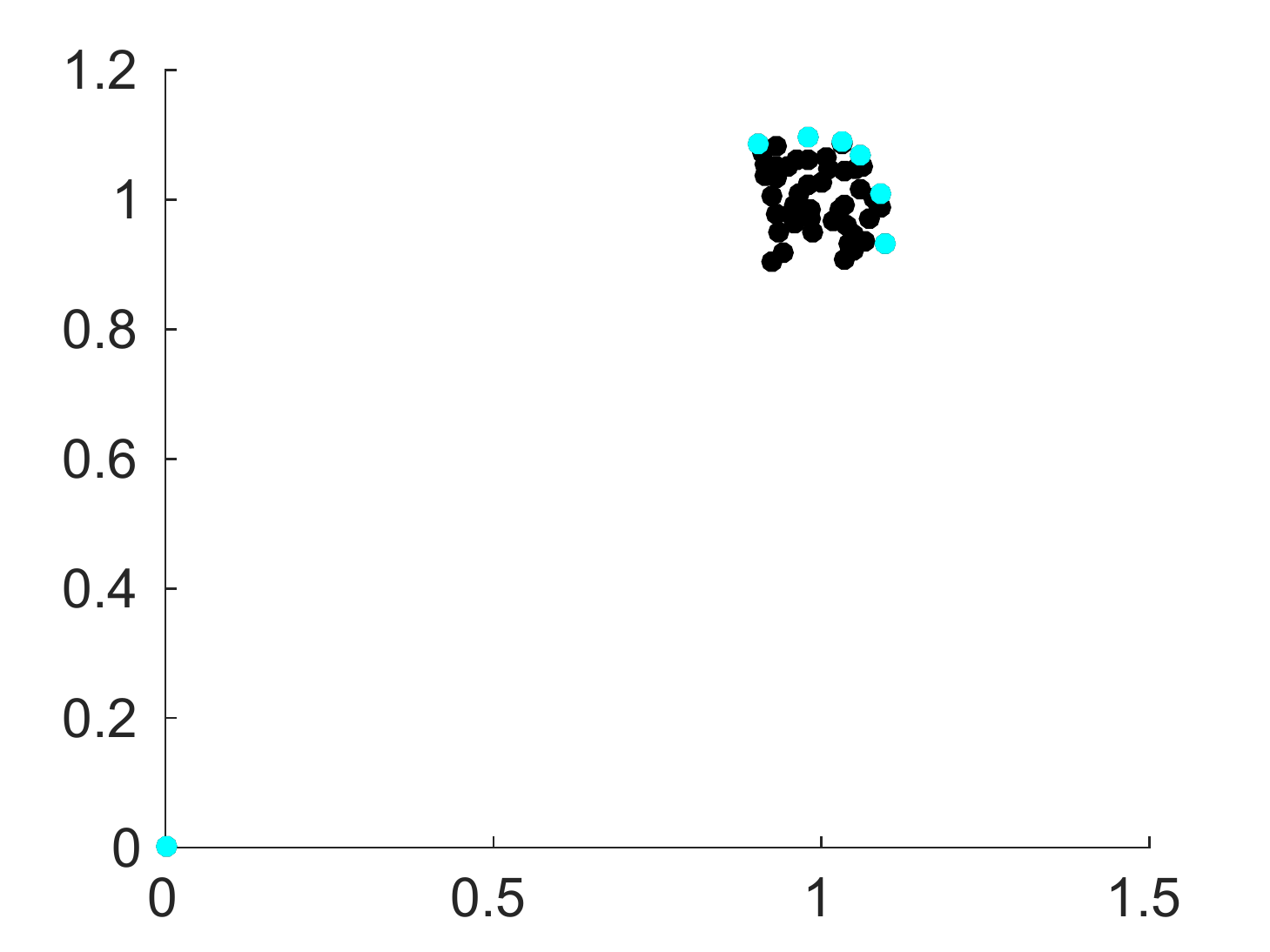}
   \includegraphics[width=0.49\textwidth]{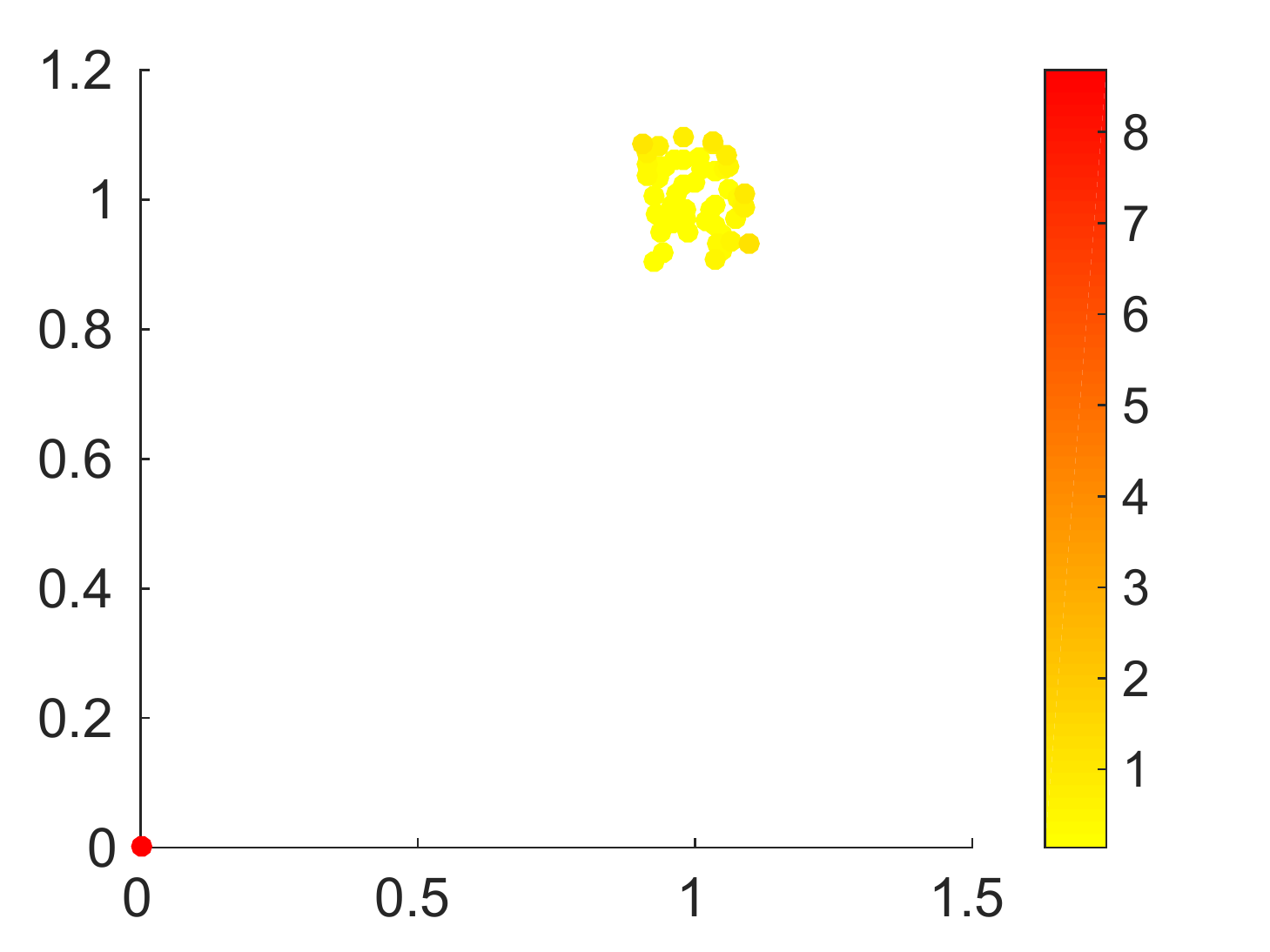}
	\caption{Distribution of points in the plane analyzed with CHSA using parameters $K=p-1$ nearest neighbors, $\lambda=10^{-3}$, and $\gamma=10^{-6}$. The left plot displays vertices of the convex hull (in cyan) with negative weights, and the right plot displays the points colored according to the norm of their weight vector.}
	\label{fig:Example2K50}
\end{figure*}

The second example is a set of data with one point at the origin and 50 others uniformly distributed  around $(1,1)$. 
\begin{figure*}
  \centering
   \includegraphics[width=0.49\textwidth]{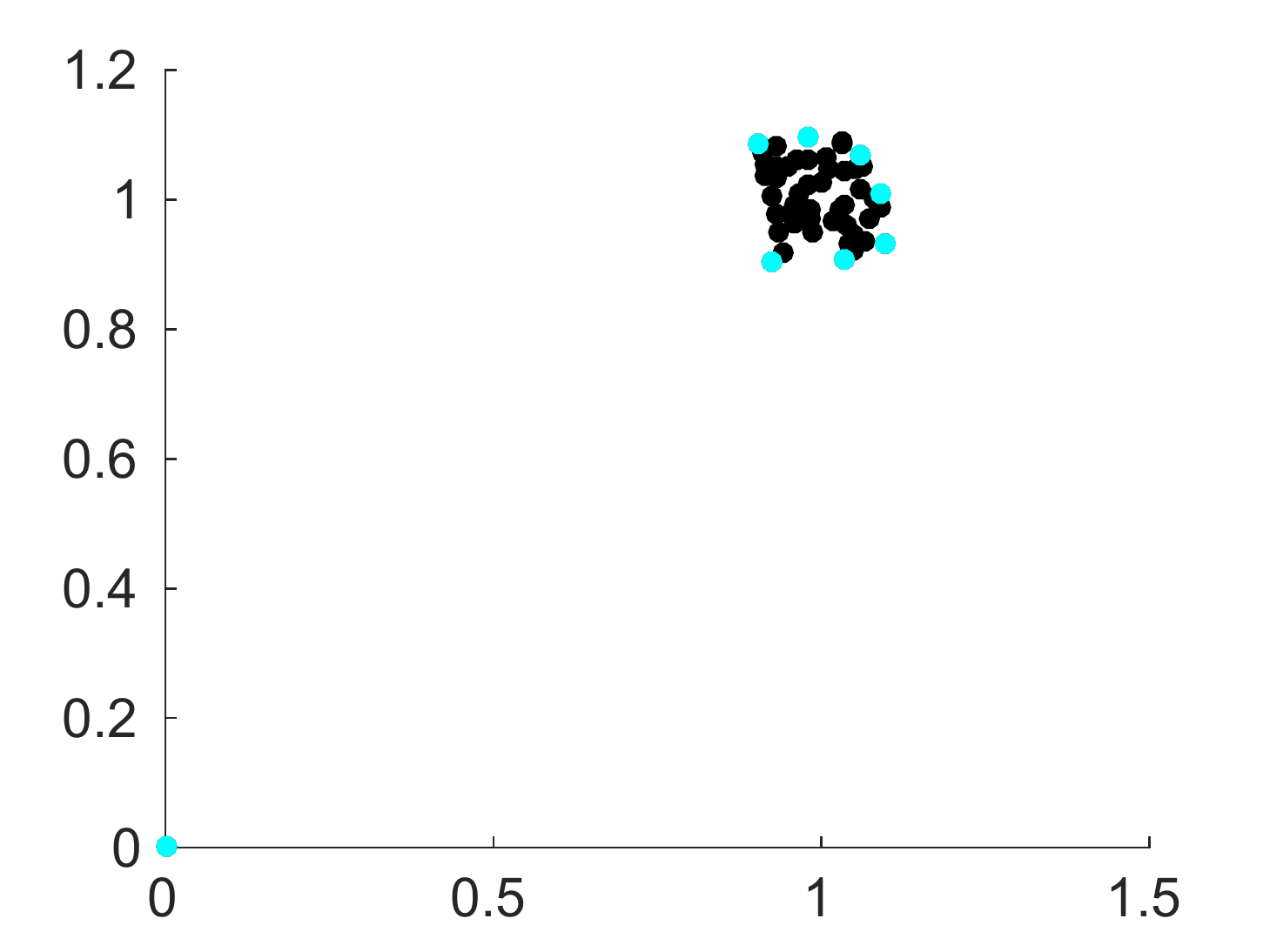}
    \includegraphics[width=0.49\textwidth]{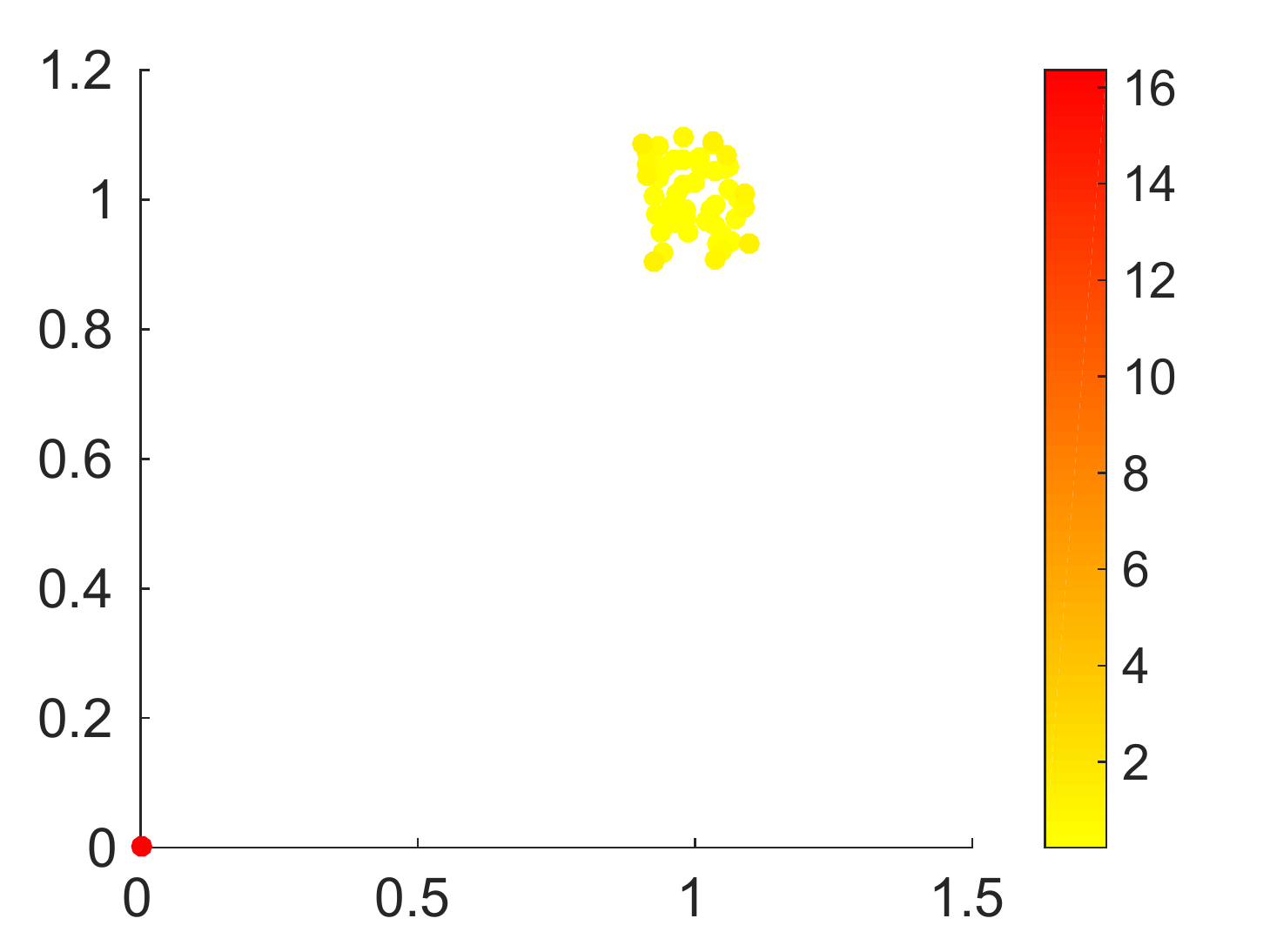}
	\caption{Distribution of points in the plane analyzed with CHSA using parameters $K=20$ nearest neighbors, $\lambda=10^{-3}$, and $\gamma=10^{-6}$. The left plot displays vertices of the convex hull (in cyan) with negative weights, and the right plot displays the points colored according to the norm of their weight vector.}
	\label{fig:Example2K20}
\end{figure*} 
In Fig. \ref{fig:Example2K20}, we have again selected parameters $K=20$ nearest neighbors, $\lambda=10^{-3}$, and $\gamma=10^{-6}$, and in Fig. \ref{fig:Example2K50}, we have modified the choice of nearest neighbors to be $K=p-1$. We observe our algorithm has similar features as in the first example, namely the norm of the weight vector is able to distinguish the outlier from all the rest, and the vertices of the interior region are revealed by having negative weights and larger magnitudes of the weight vector than those points inside the interior region.

\begin{figure*}
  \centering
  \includegraphics[width=0.49\textwidth]{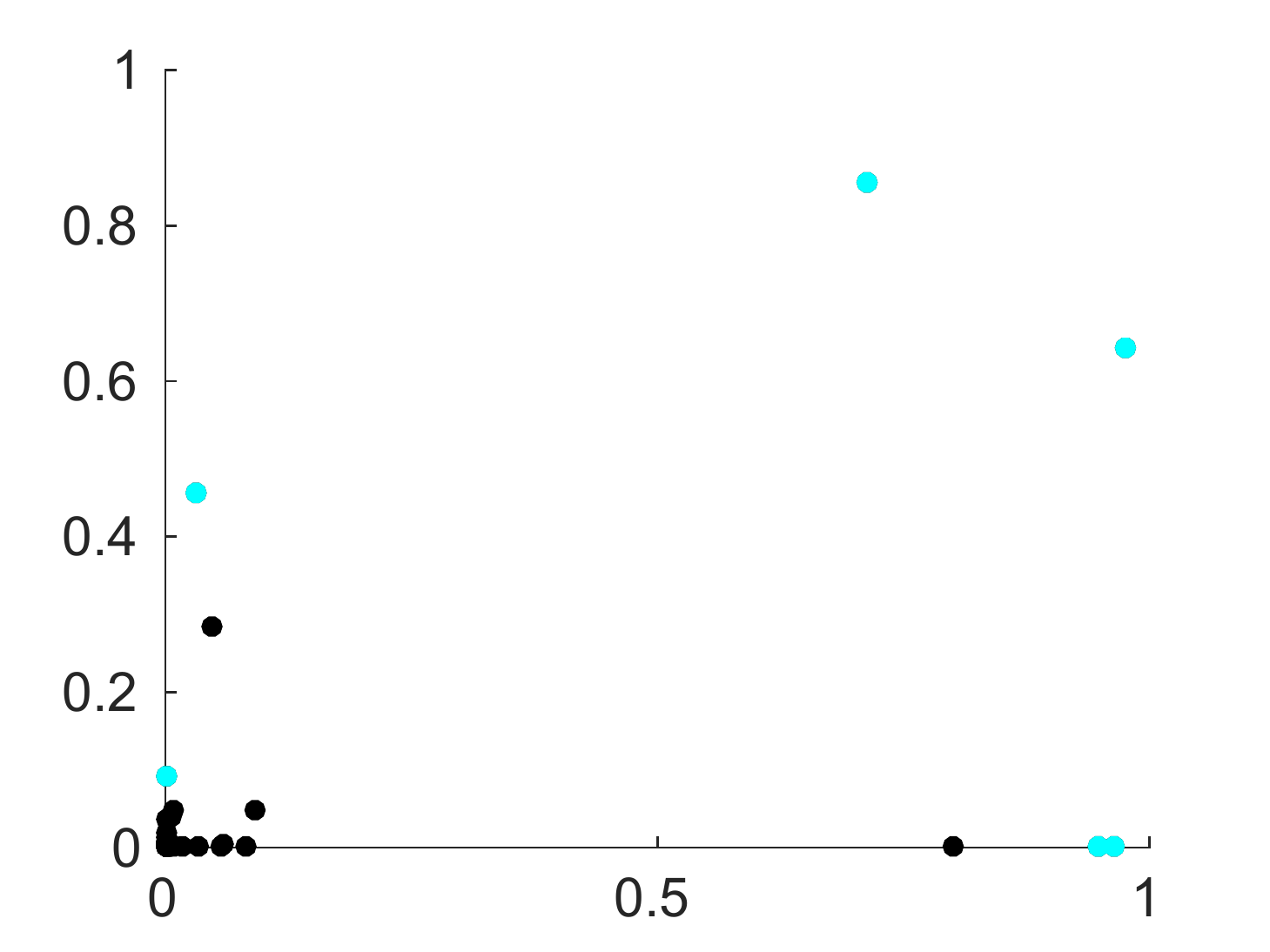}
   \includegraphics[width=0.49\textwidth]{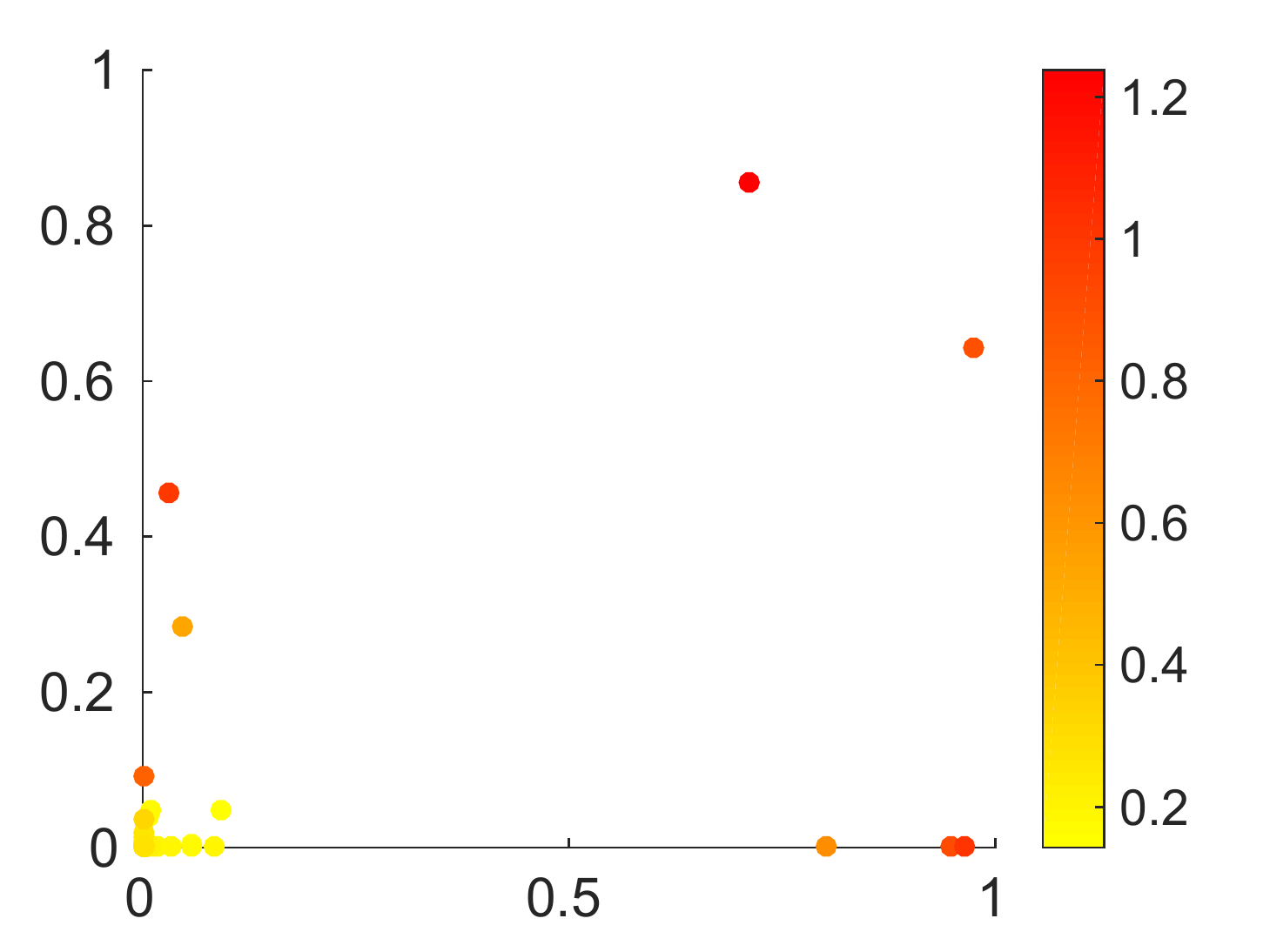}
	\caption{Logarithmic distribution of points in the plane analyzed with CHSA using parameters $K=p-1$ nearest neighbors, $\lambda=10^{-3}$, and $\gamma=10^{-6}$. The left plot displays vertices of the convex hull (in cyan) with negative weights, and the right plot displays the points colored according to the norm of their weight vector.}
	\label{fig:Example3K49}
\end{figure*}

\begin{figure*}
  \centering
  \includegraphics[width=0.49\textwidth]{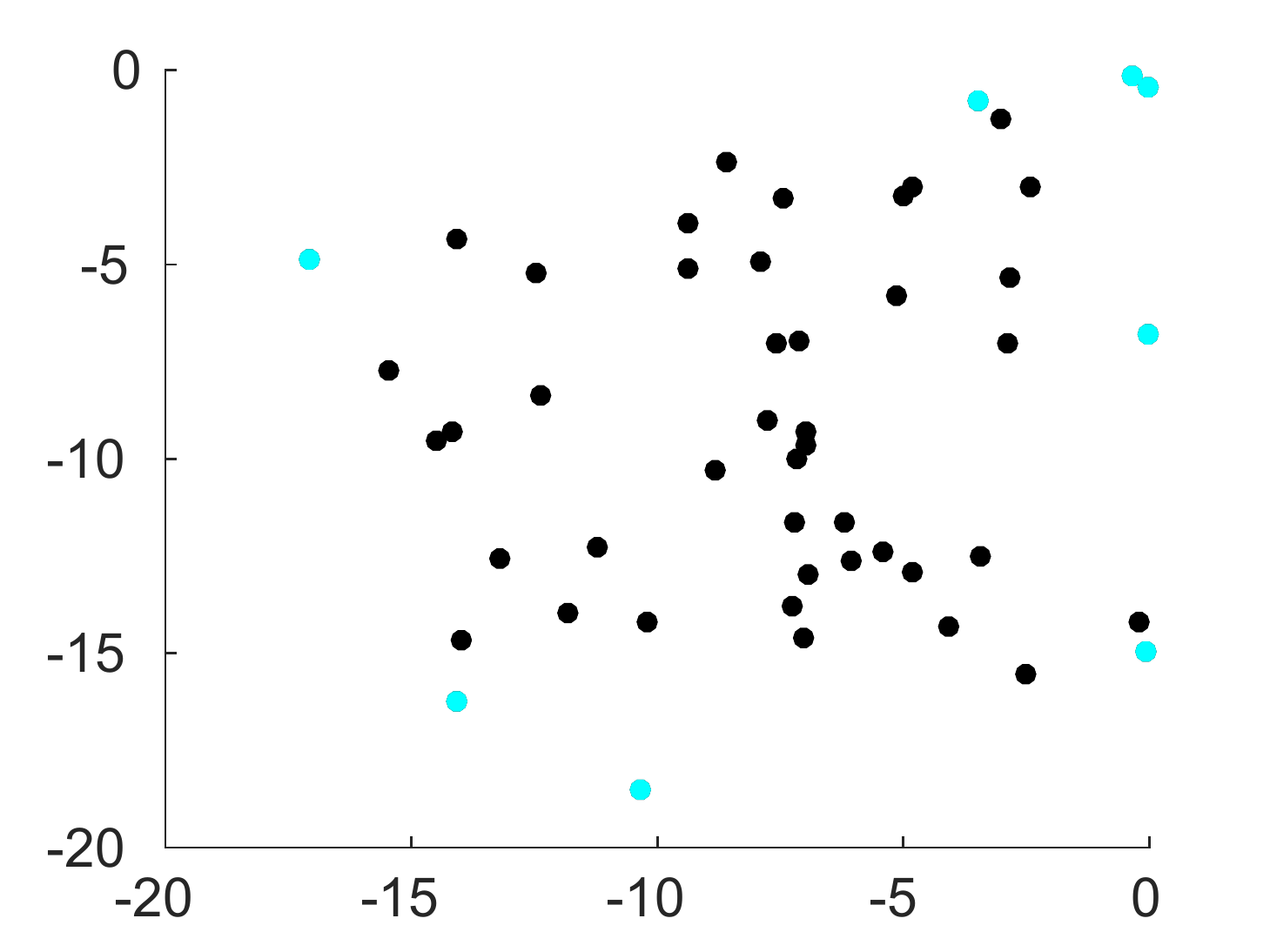}
 \includegraphics[width=0.49\textwidth]{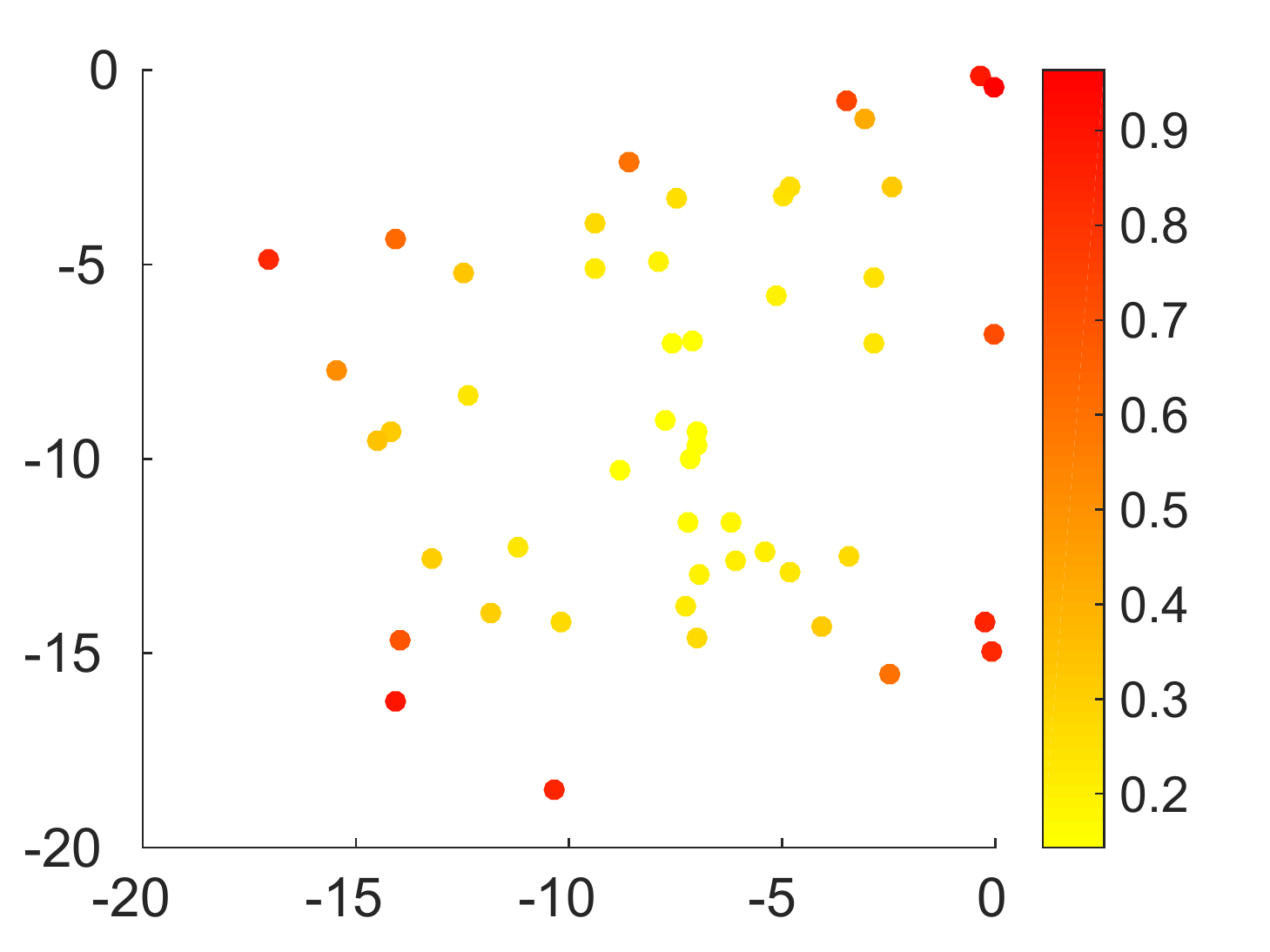}
	\caption{Logarithmic distribution of points in the plane pre-processed by taking the logarithm and analyzed with CHSA using parameters $K=p-1$ nearest neighbors, $\lambda=10^{-3}$, and $\gamma=10^{-6}$. The left plot displays vertices of the convex hull (in cyan) with negative weights, and the right plot displays the points colored according to the norm of their weight vector.}
	\label{fig:Example3K49Log}
\end{figure*}

In the final example, we have constructed logistic random data in the range $[10^{-8},1]$ and implemented CHSA with parameters $K=p-1$ nearest neighbors, $\lambda=10^{-3}$, and $\gamma=10^{-6}$. Results are displayed in Fig. \ref{fig:Example3K49}. We observe that while our algorithm does not capture all of the vertices of the convex hull (particularly those small in magnitude near the origin), it does capture most of them, and again the norm leads to a useful stratification of the data. As a bit of standard pre-processing that would be reasonable with this sort of data, we take the logarithm and implement CHSA with the same choice of parameters. Results are displayed in Fig. \ref{fig:Example3K49Log}. We observe that with this transformed data our algorithm does in fact uncover the correct vertices of the convex hull, and again, the norm reveals a useful stratification of the data. 

To summarize this section, we highlight some of the benefits and drawbacks of CHSA. If one wishes to have only the vertices of the convex hull uncovered as having negative weights, a choice of $K=p-1$ nearest neighbors will do so, but this is computationally expensive. However, if one wants to use a smaller choice of nearest neighbors to speed up computations, CHSA still reveals useful (perhaps even more useful) information. That is, it not only reveals the true vertices of the convex hull as having negative weights but also candidate vertices of the interior cluster. This, in conjunction with the norm of the weight vector, can still reveal the true vertices but also uncovers more of the structure of the entire data set.

\subsection{Convex hull detection of cube}

\begin{figure}
\centering
\includegraphics[width=.6\textwidth]{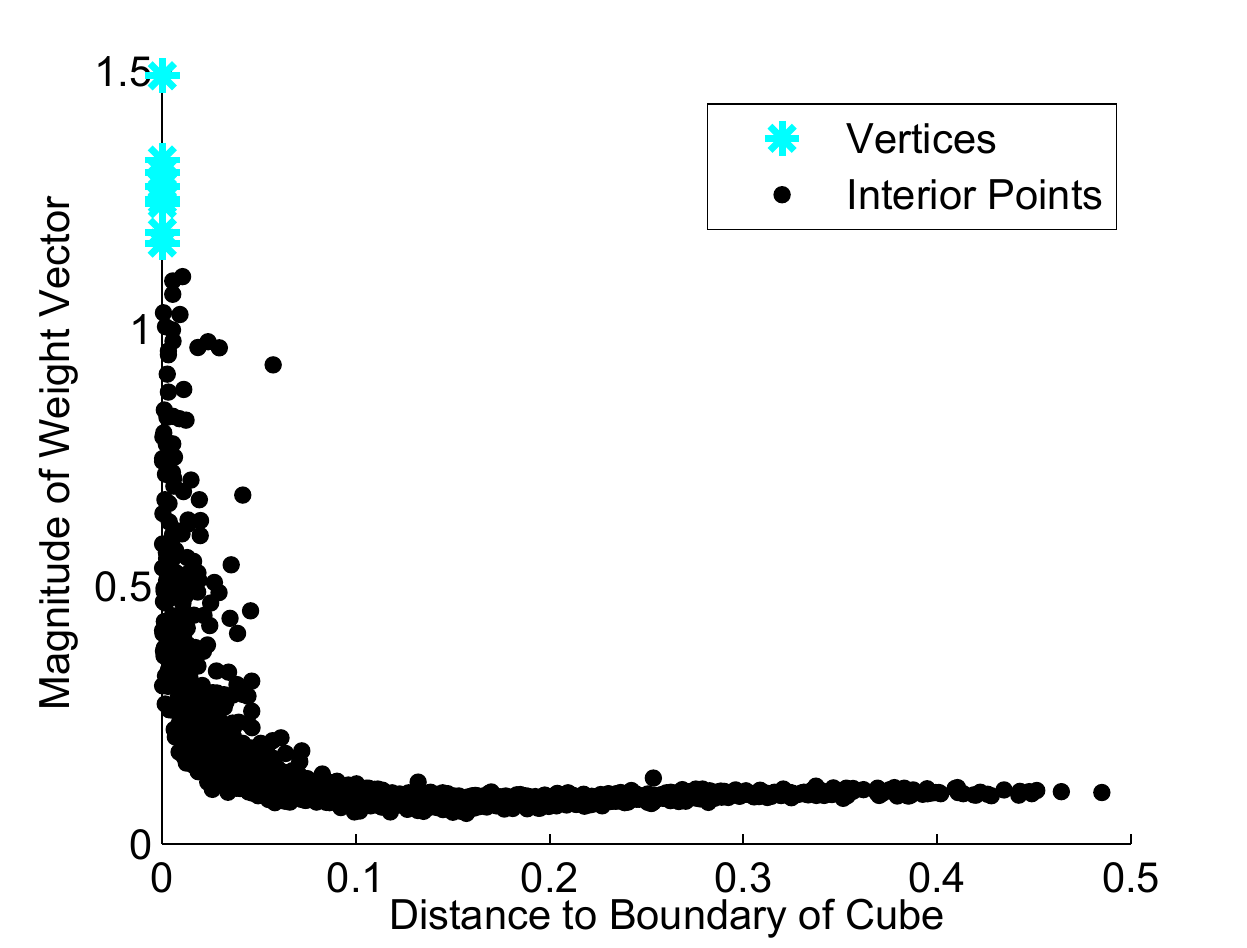}
\caption{Implementation of CHSA on 2000 randomly distributed points inside the unit cube together with the 8 vertices of the cube with parameters $\gamma=10^{-5}$ and $\lambda=0.025$. Plot of the distance to the boundary of the cube versus the Euclidean norm of weights associated to each point.}
\label{fig:NormsDistance} 
  \end{figure}

In the next example, we consider 2000 uniformly distributed random points inside the unit cube together with the 8 vertices of the cube.  We choose $K=200$ nearest neighbors to represent each point, fix $\gamma=10^{-5}$, and explore the effect of varying $\lambda$ in CHSA.  For $\lambda=0.001$, we find 40 points with at least one negative weight falling very close to the vertices, edges, and faces of the cube.  As $\lambda$ increases, the number of points with negative weights decreases. When $\lambda=0.005$,  18 points survive whose weight vector contains negative weights. When $\lambda=0.01$, there are 12 points whose weight vector contains negative weights. When $\lambda=0.025$, only the 8 vertices of the cube have negative weights. In this last setting, 
we consider the Euclidean norms of the weight vectors associated to each of the 2000 points, see Fig. \ref{fig:NormsDistance}.  The 8 vertices of the cube correspond to the largest Euclidean norms.  Next, we find points close to the vertices, then close to the edges, then close to the faces of the cube. The majority of the points, which are not close to the boundary of the cube, are not distinguished by their Euclidean norm.

\subsection{Convex combination of chemical signatures}

In Section \ref{firsteg}, we introduced the interpretation of vertices of the convex hull as being special in the sense that they can be viewed as the ingredients for all other pixels in the interior of the convex hull.  In other words, any point $\textbf{x} \in C$ can be expressed as the superposition of the vertices $\{\textbf{v}_i\}$ \textit{i.e.} 
\begin{equation}
\label{decomp}
\textbf{x} = \sum_{i=1}^N a_i \textbf{v}_i
\end{equation}
where $a_i$ is viewed as the {\it abundance}  of  the pure substance $\textbf{v}_i$. This decomposition of a point into its pure components is at the heart of the analysis of hyperspectral imagery and illustrates the importance of being able to compute the vertex set $\{\textbf{v}_i\}$. The number of vertices $N$ determined by the algorithm provides critical insight into the number of pure substances that are present in the field of view of the hyperspectral image \cite{greer2012sparse,zare2007sparsity}.

\begin{figure}
  \centering
	\hspace{-1cm}
	\includegraphics[width=.6\textwidth]{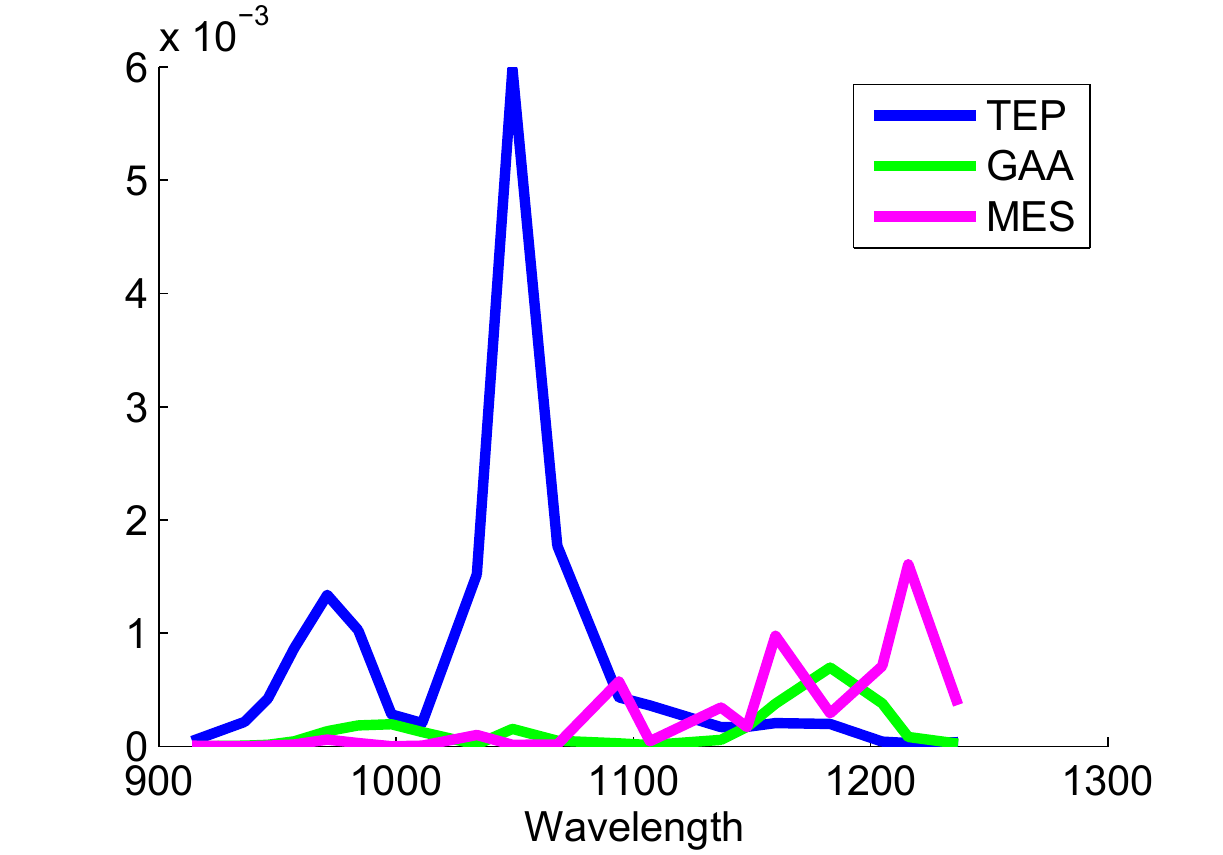}
  \caption{Spectral signatures of the chemical simulants Glacial Acetic Acid (GAA), Methyl Salicylate (MeS), and Triethyl Phosphate (TEP) as a function of wavelength represented in $\mathbb{R}^{20}$.}
  \label{fig:ConvHull20DSignatures}
\end{figure}

To illustrate this, we have chosen to analyze data produced by weighted mixtures of pure chemical simulants lying in a high dimensional spectral space.  For simplicity we consider only three pure substances \textit{i.e.}  Glacial Acetic Acid (GAA), Methyl Salicylate (MeS), and Triethyl Phosphate (TEP), each represented by a 20-dimensional spectral signature as a function of wavelength, see Fig. \ref{fig:ConvHull20DSignatures}.  
Given these three vertices $\{ \textbf{v}_i \}$ of a convex hull in $\mathbb R^{20}$, we generate 1000 additional sample combinations using Equation (\ref{decomp}) and uniformly distributed random $\{a_i\}$.  This produces a 2-dimensional convex set of mixed simulants in 20 dimensions whose vertices are the pure chemicals. As can be seen in Fig. \ref{fig:ConvHull20DSignatures}, this data is between 0 and $6 \times 10^{-3}$; we rescale to be between 0 and 1.
 
\begin{figure*}
  \centering
  \subfloat[$\lambda=10^{-7}$, $\gamma=10^{-6}$]{\label{fig:CC1} \includegraphics[width=.5\textwidth]{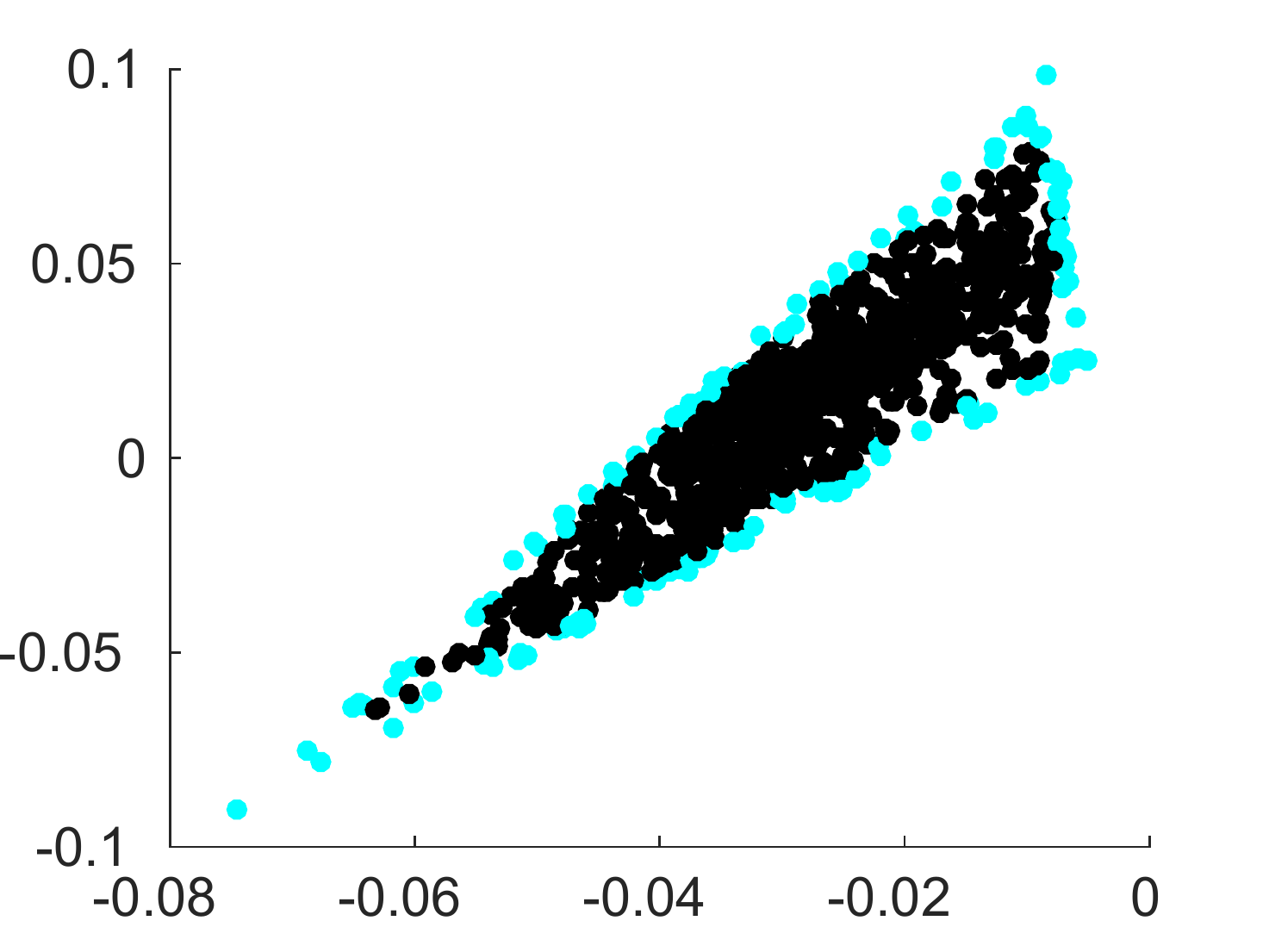}}
  \subfloat[$\lambda=10^{-5}$, $\gamma=10^{-6}$]{\label{fig:CC2} \includegraphics[width=.5\textwidth]{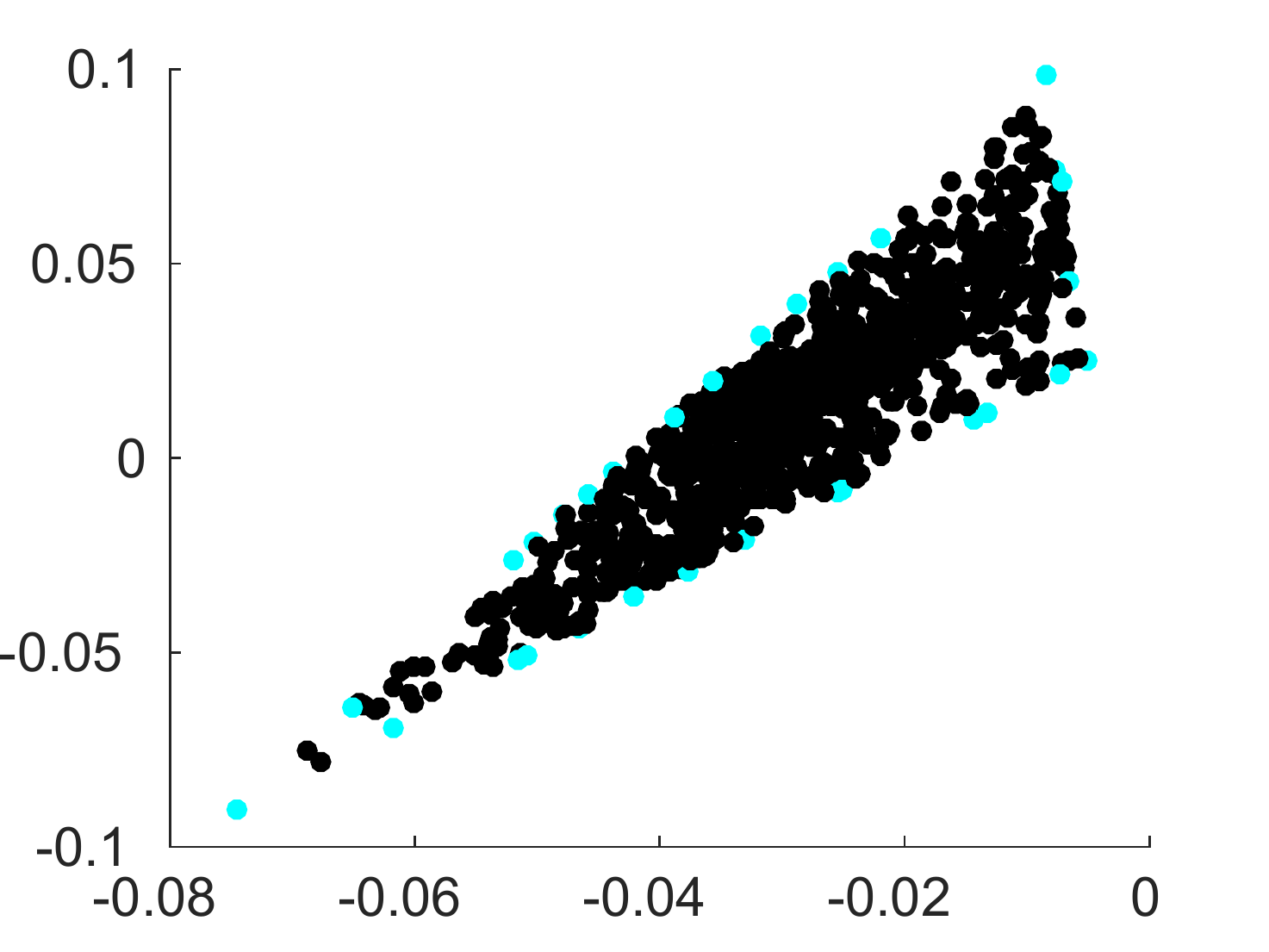}}\\
   \subfloat[$\lambda=10^{-3}$, $\gamma=10^{-6}$]{\label{fig:CC3} \includegraphics[width=.5\textwidth]{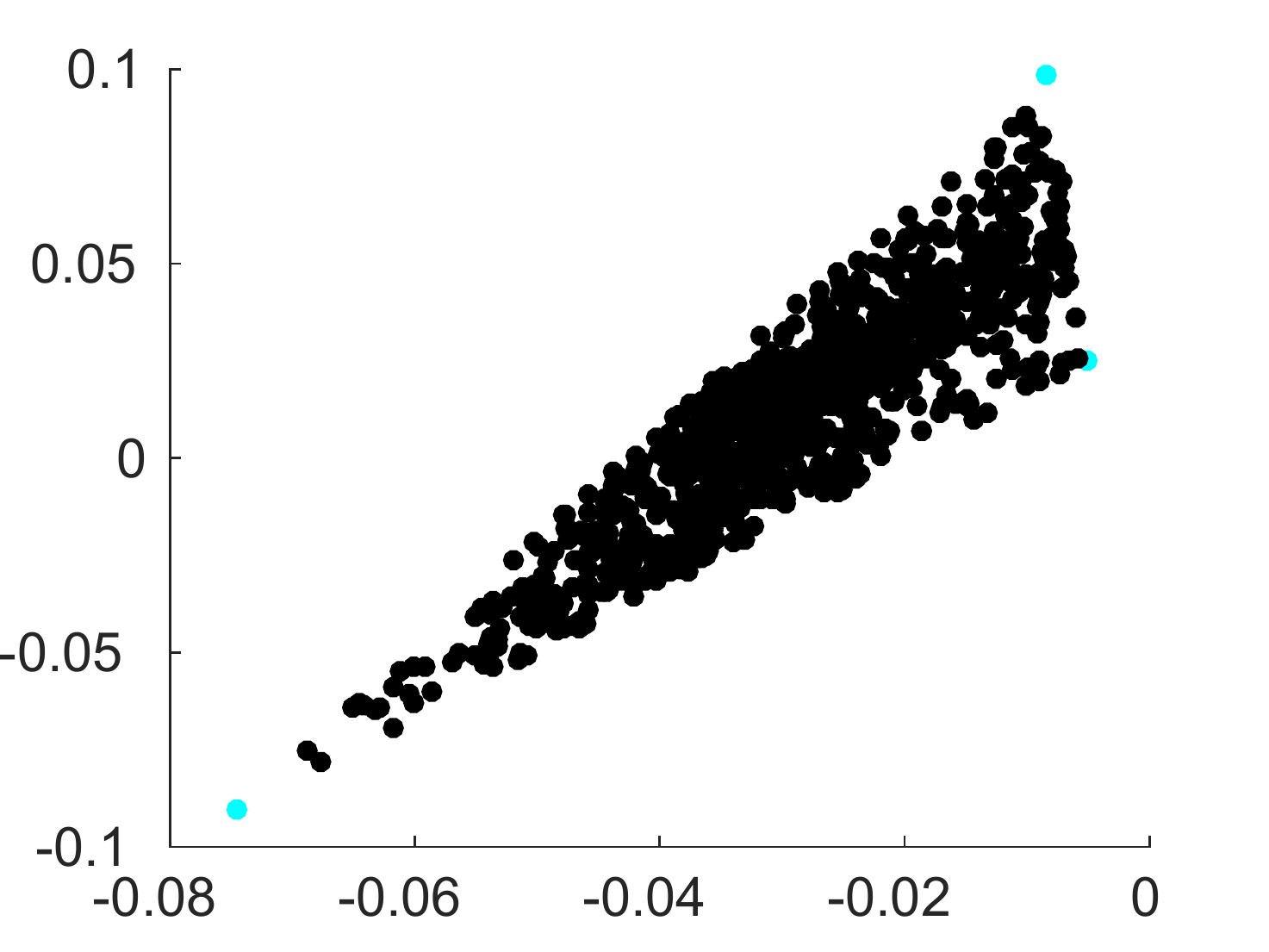}}
	\subfloat[$\ell_2$ Norms of Weights]{\label{fig:CCMagWeight} \includegraphics[width=.5\textwidth]{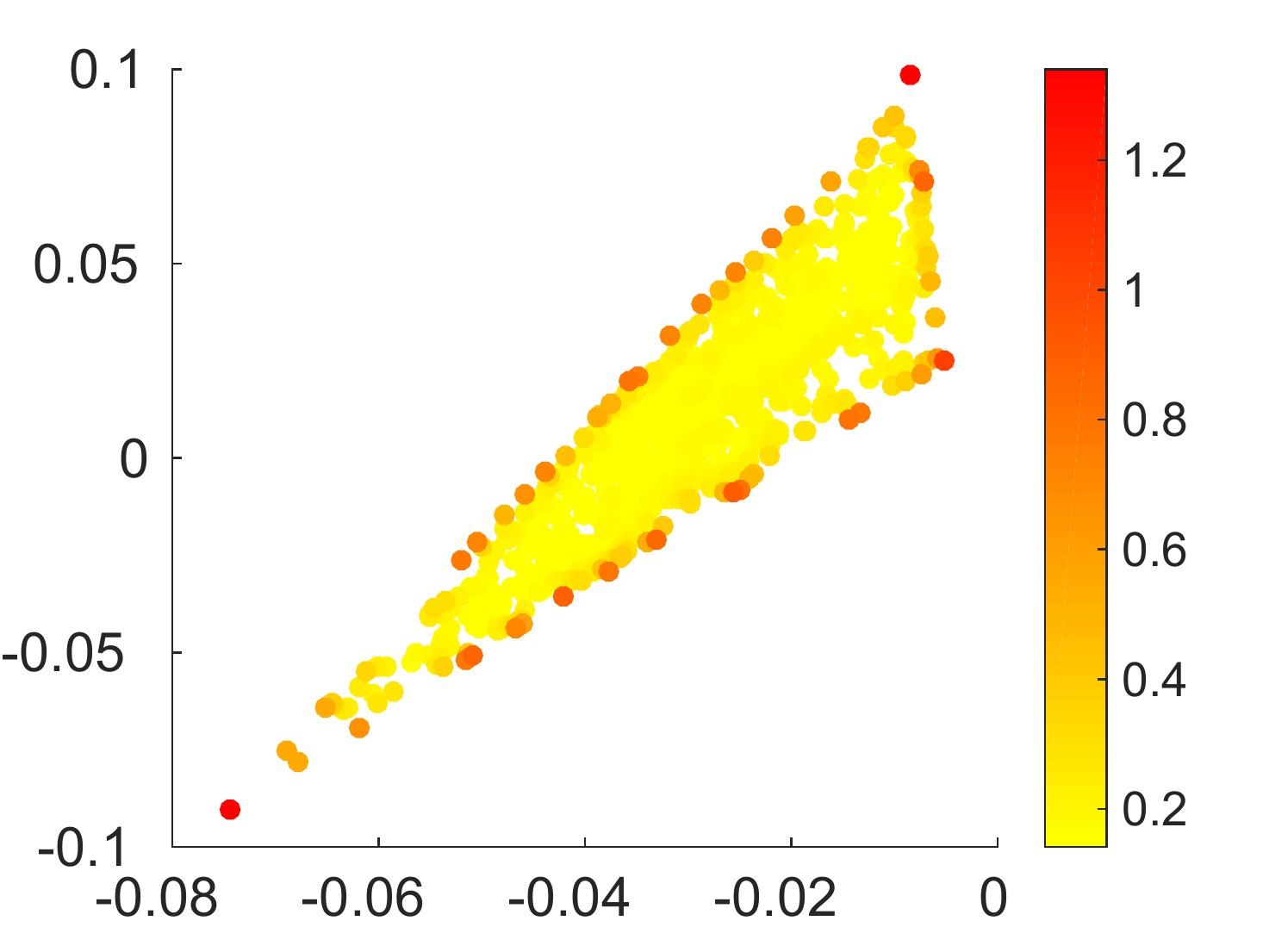}}
  \caption{(a)-(c) Implementation of CHSA with various parameter choices on convex combination created using 20 dimensional chemical signatures, projected down to $\mathbb{R}^2$ using PCA.  Axes of the plots are the coordinates of the first two principal vectors. Points represented in cyan have a negative weight. (d) Data points colored according to the magnitude of the $\ell_2$ norm (using parameters $\lambda=10^{-3}, \gamma=10^{-6}$) where yellow corresponds to the smallest values and red corresponds to the largest values.}
  \label{fig:ConvHullDTRA}
\end{figure*}

Implementations of CHSA reveal the structure of this data. We fix $K=50$ nearest neighbors and $\gamma=10^{-6}$, and 
explore the effect of varying $\lambda$.  Using principal component analysis (PCA) \cite{jolliffe2002principal}, we project the data set into $\mathbb{R}^2$ for visualization purposes.  Once again, cyan indicates those data points whose weight 
vector contains negative weights (see Fig. \ref{fig:ConvHullDTRA}). The axes of these plots are the coordinates of the two principal vectors. 
For a choice of $\lambda=10^{-7}$, we see a thick cyan boundary, with $\lambda=10^{-5}$ we see a thinner cyan boundary, and with $\lambda=10^{-3}$ we see only the vertices of the triangle.  Thus, not only are the vertices, or pure substances, recovered by the algorithm, the structure of the data is revealed at multiple scales.  In Fig. \ref{fig:CCMagWeight}, we color each of the points according to the $\ell_2$ magnitude of the weight vector (from optimizing with parameters $\lambda=10^{-3}$ and $\gamma=10^{-6}$), which reveals a point's proximity to the boundary.  This leads to a coarse stratification by pureness of each point.

Notice the behavior with respect to choices of parameters $\lambda$ and $\gamma$ of this experiment is similar to the experiments previously discussed despite having points in a higher ambient dimension. 

\subsection{Return to the sand example}

\begin{figure}
\centering
\subfloat[]{\label{fig:Sand1Norms}\includegraphics[width=.5\textwidth]{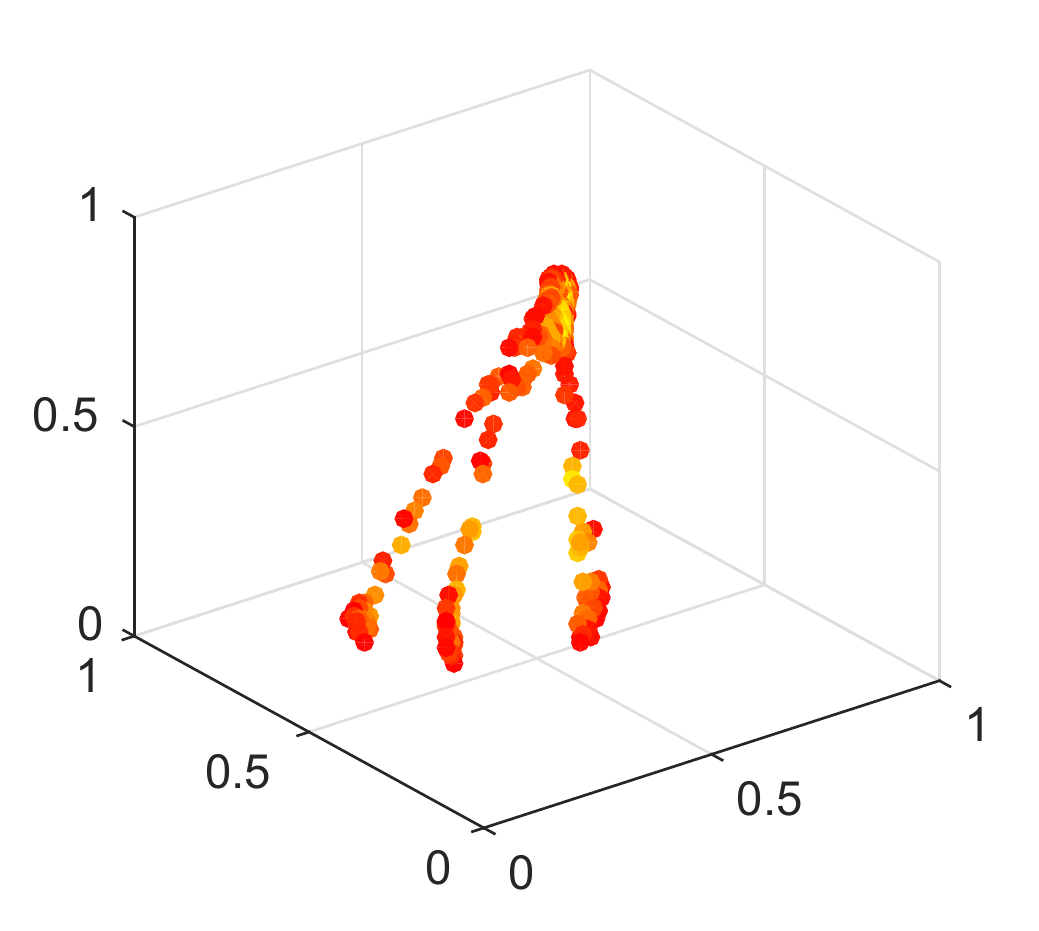}}
\subfloat[]{\label{fig:Sand2Norms}\includegraphics[width=.5\textwidth]{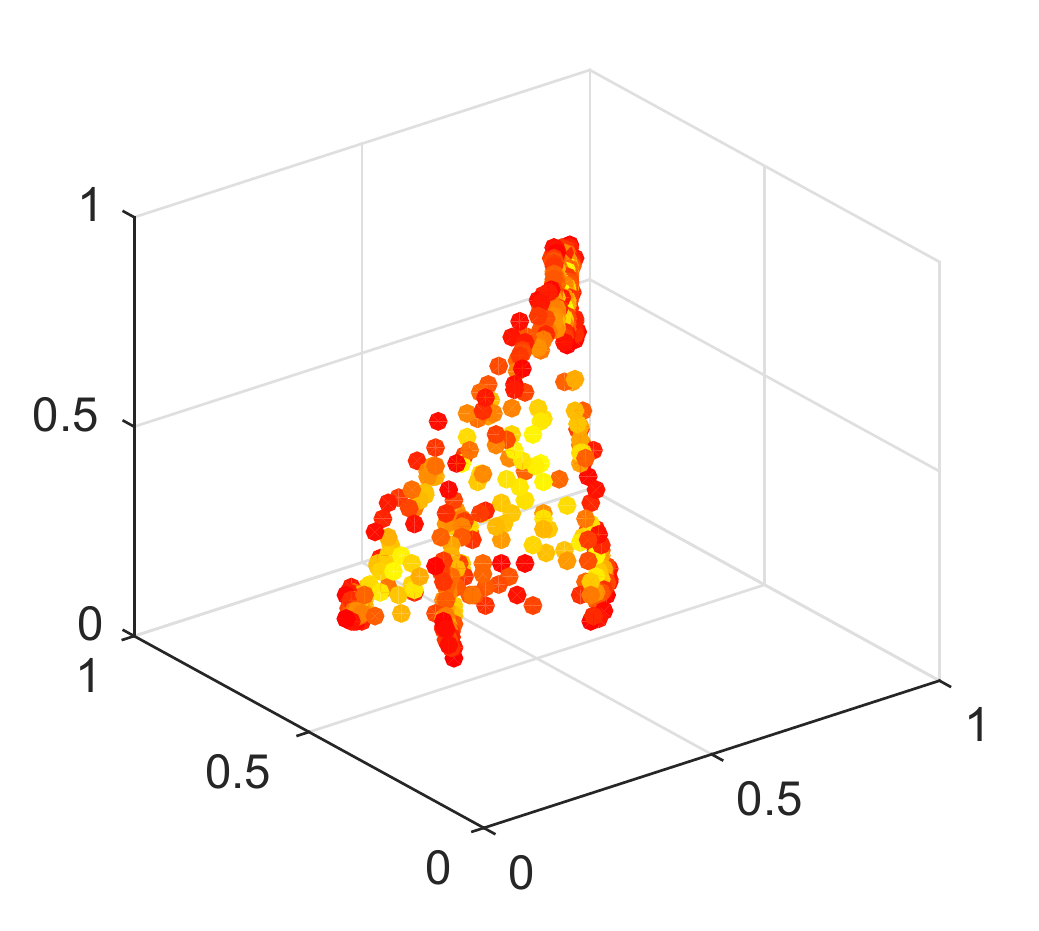}}\\
\subfloat[]{\label{fig:Sand3Norms}\includegraphics[width=.5\textwidth]{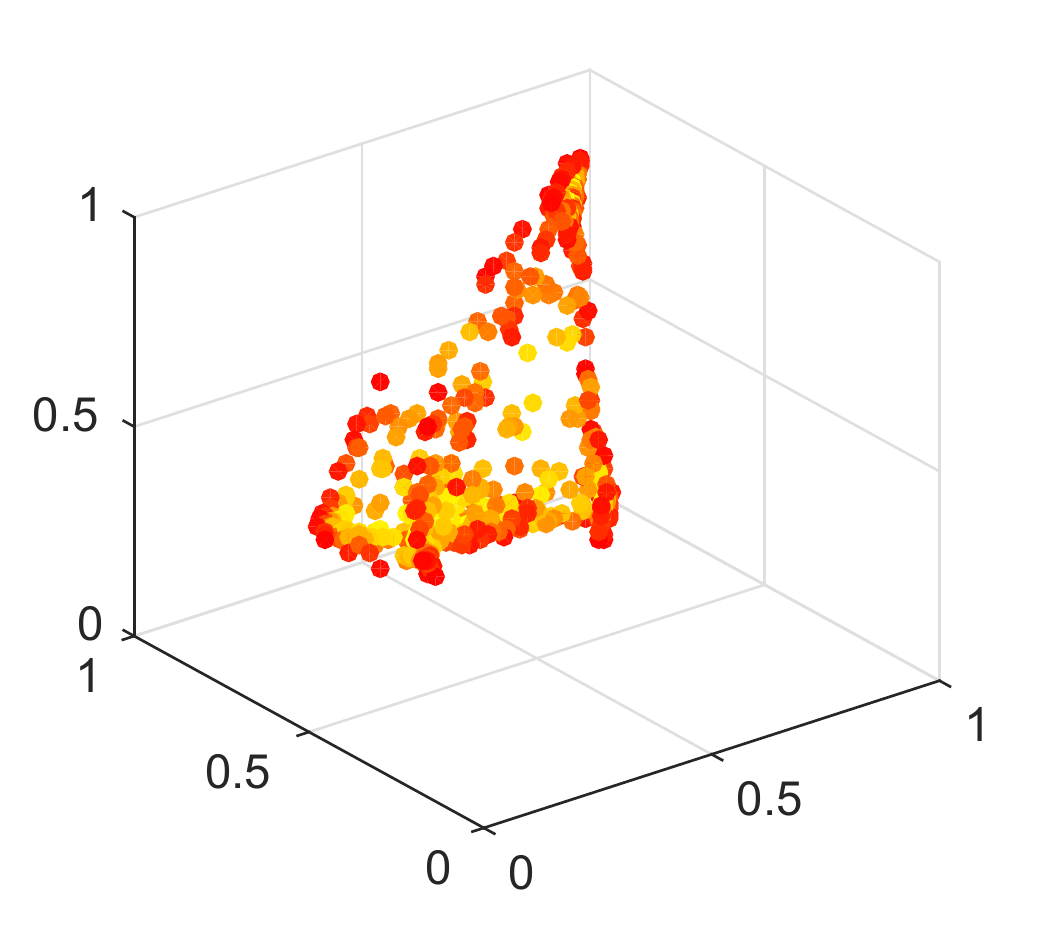}}
\subfloat[]{\label{fig:Sand4Norms}\includegraphics[width=.5\textwidth]{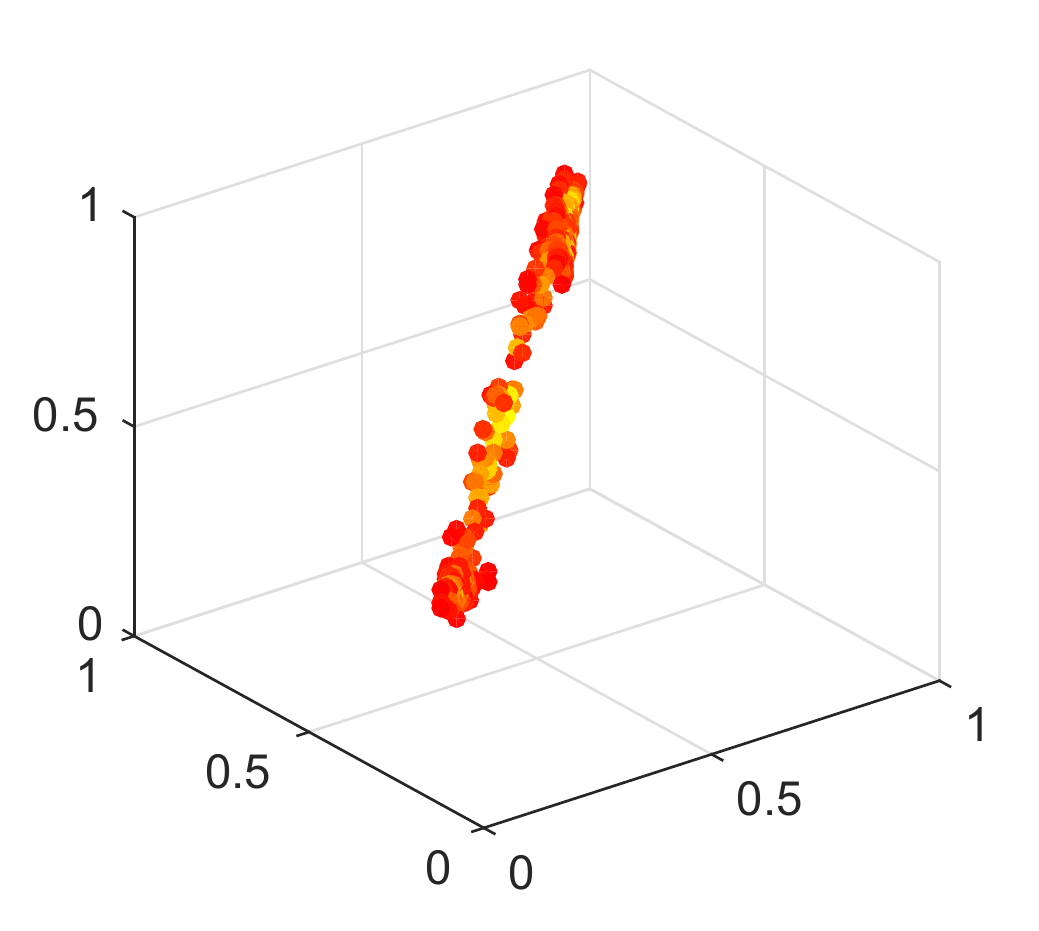}}
\caption{Data points of sand images colored according to ordering of the norms of the $\ell_2$ norm where yellow corresponds to the smallest values and red corresponds to the largest values, using parameters $K=100, \lambda=10^{-3}, \gamma=10^{-6}$.}
\label{fig:SandNorms}
\end{figure}

Finally, we return to our motivation application discussed in Section \ref{firsteg}, an experiment of three colors of sand poured onto a sheet of paper and mixed together at different levels.  We solve the optimization problem with parameters $K=100, \lambda=10^{-3}, \gamma=10^{-6}$, and then consider the structure revealed by the weight vector corresponding to each point. We color each data point according to the magnitude of the weight vector in Fig. \ref{fig:SandNorms} slightly differently than before for better visualization. In this instance, the color corresponds to the ordering of the norms of the weight vectors instead of the actual magnitudes of the norms of the weight vectors. We again observe that CHSA stratifies points along the vertices, boundaries, and interior.

\begin{figure}
\centering
\subfloat[]{\label{fig:LargestNorms}\includegraphics[width=.33\textwidth]{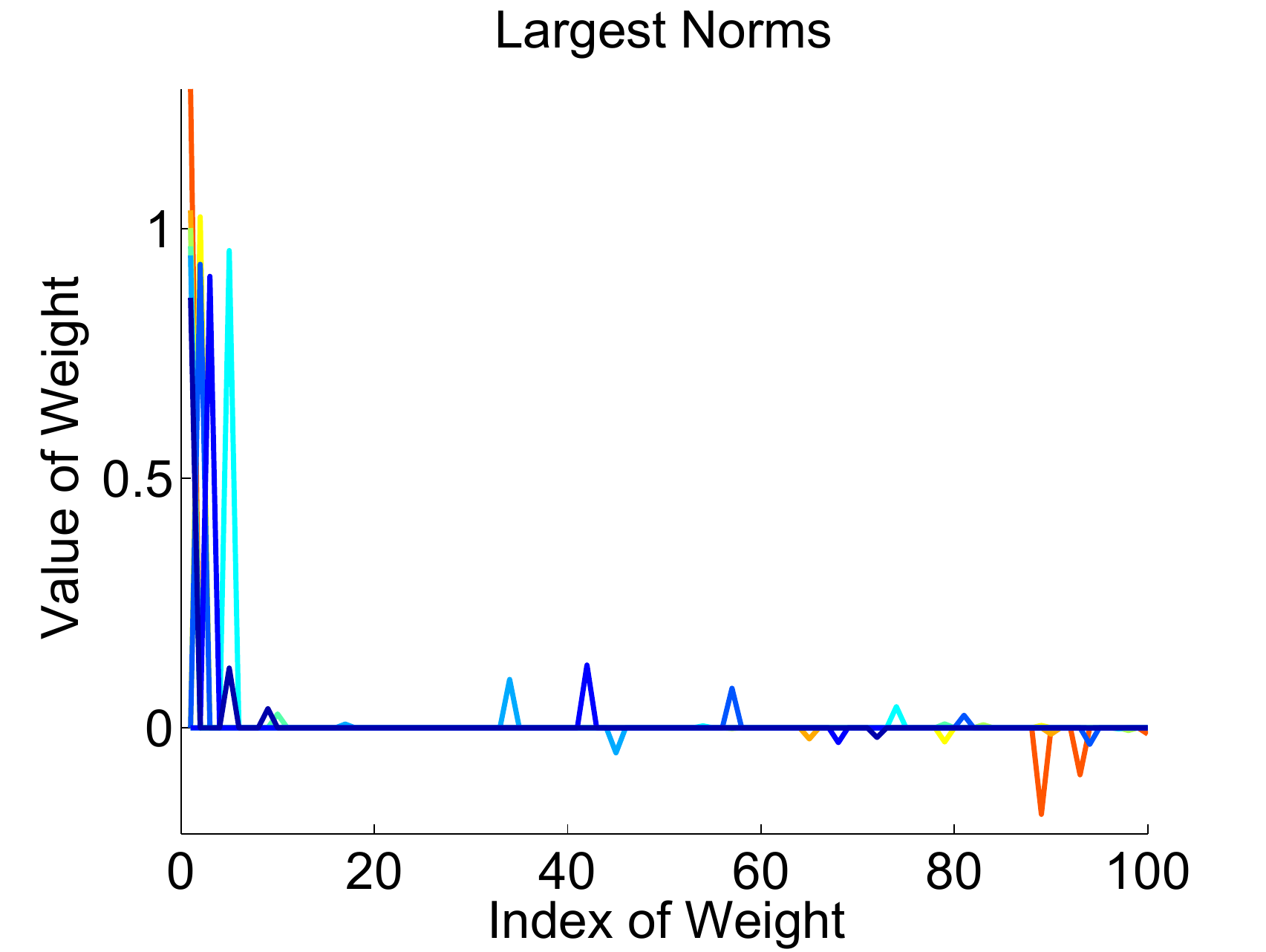}}
\subfloat[]{\label{fig:MiddleNorms}\includegraphics[width=.33\textwidth]{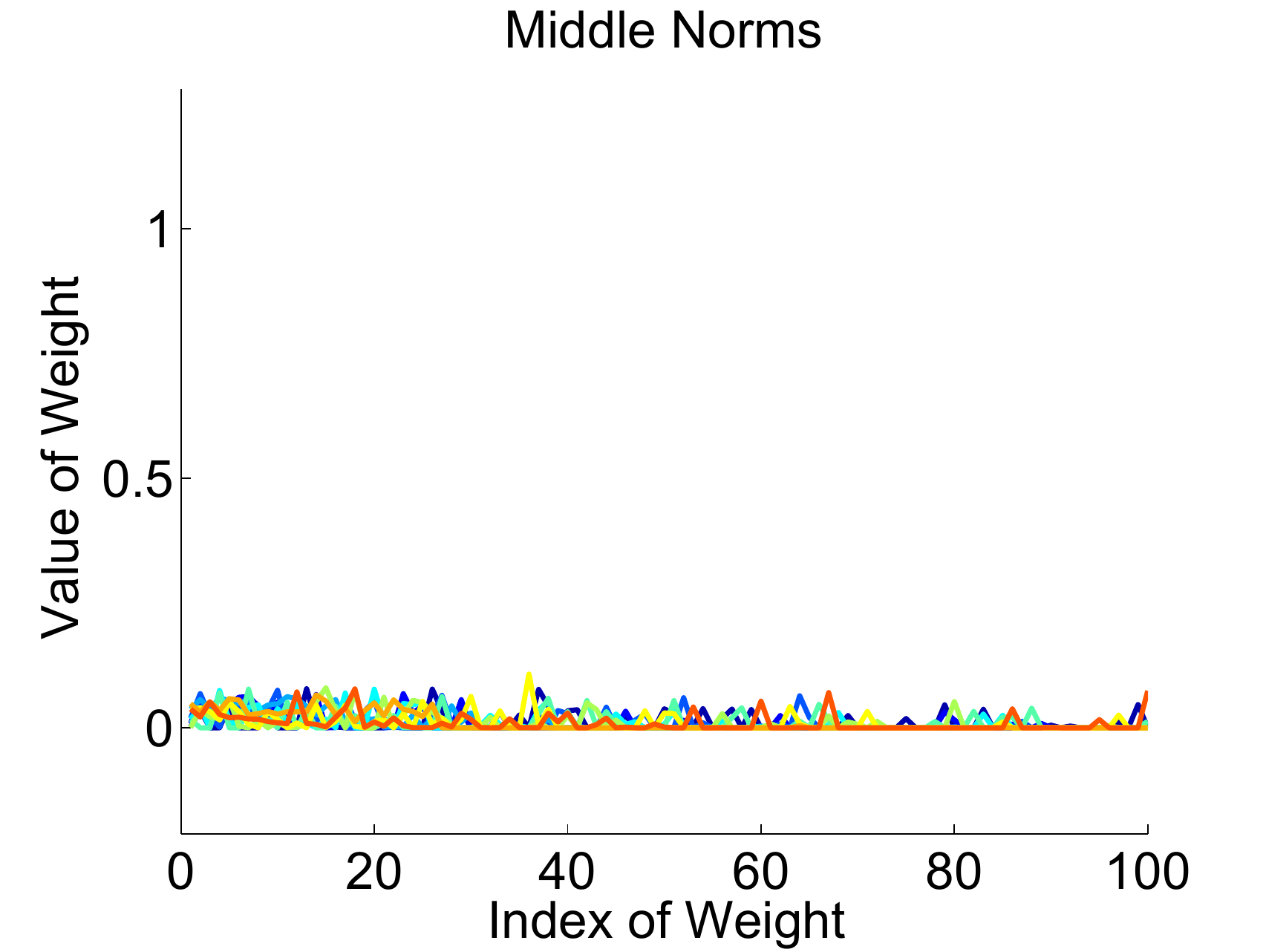}}
\subfloat[]{\label{fig:SmallestNorms}\includegraphics[width=.33\textwidth]{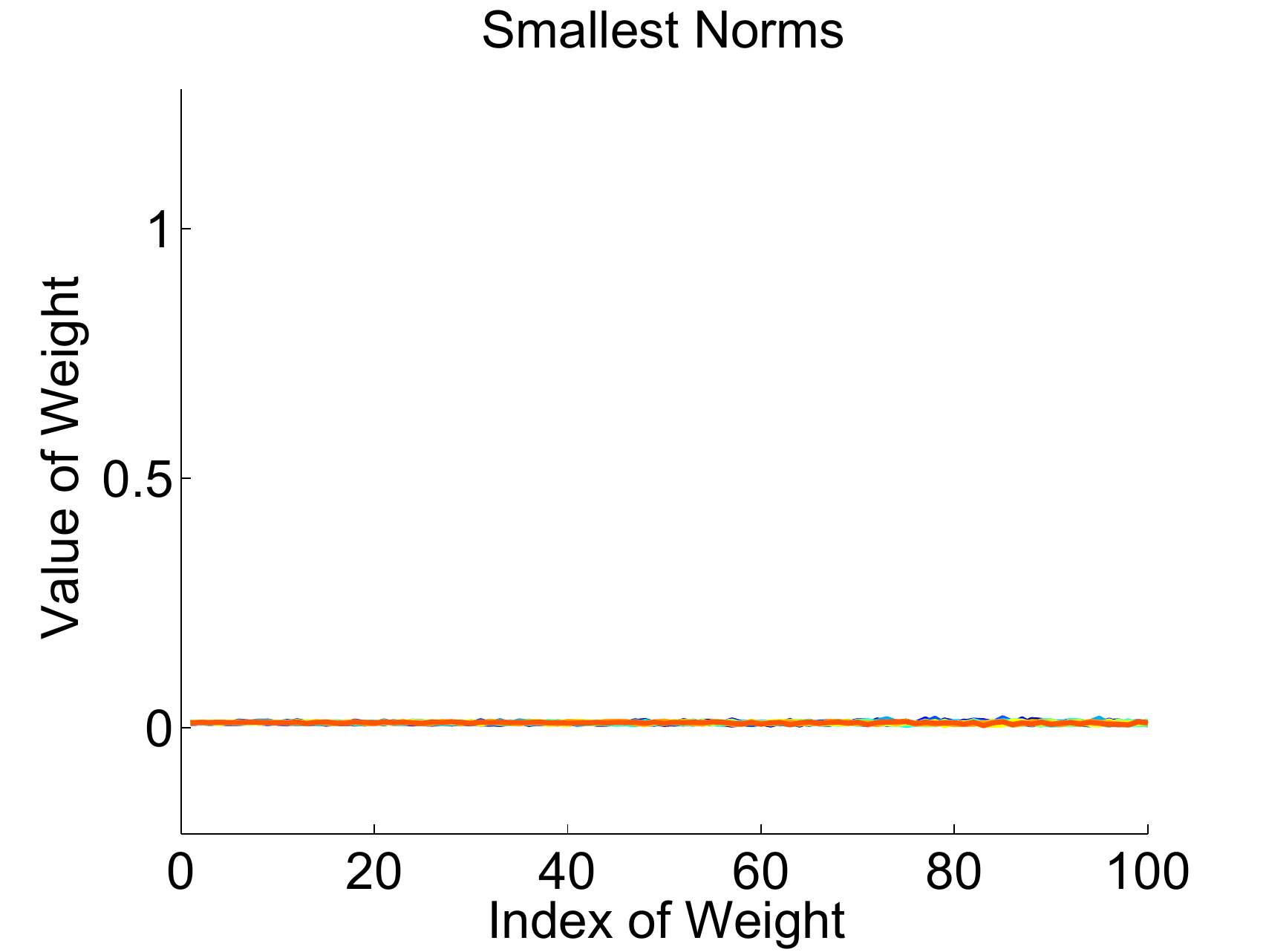}}
\caption{Plot of the weight vectors associated to points from the first sand image with the 10 largest Euclidean norms, 10 surrounding and including the median, and 10 smallest Euclidean norms.}
\label{fig:SandWeightNorms}
\end{figure}

As final evidence that the weight vectors (and thus, the norms of these vectors) are quantitatively different, we sort the weight vectors according to their norms and plot the 10 weight vectors from the first image with the largest norms, the 10 weight vectors surrounding and including the median of the norms (which we refer to henceforth as the middle norms), and the 10 weight vectors with the smallest norms in Fig. \ref{fig:SandWeightNorms}.
Let us now make three observations.  First, there are no negative weights in the set of vectors corresponding to the middle and smallest norms, yet there are negative weights in the set of vectors corresponding to the largest norms, indicating vertices of the convex hull.  Second, the vectors corresponding to the middle and smallest norms are much more uniform than those corresponding to the largest norms, with the components of the weight vectors associated to the smallest norms all being approximately $1/K$.  Finally, sparsity is induced in the weight vectors corresponding to the largest norms as a consequence of the $\ell_1$ term in the objective function.  Note that for points with no negative weights this term is essentially ignored.

\section{Conclusion}
We have proposed an optimization problem that can be used to identify vertices of the convex hull of a data set in high dimensions and to stratify points in the data set based on proximity to the convex hull which we call the Convex Hull Stratification Algorithm (CHSA).  The optimization problem consists of three components.   Primary emphasis is placed on the term concerned with representing a point as an affine combination of its neighbors with minimal error.   The other two components, $\ell_1$ and $\ell_2$ penalty terms, are used to encode geometric structure into the weight vector $\textbf{w}$ associated with a data point $\textbf{x}$.   The $\ell_1$ and $\ell_2$ penalty terms are weighted by parameters $\lambda$ and $\gamma$, respectively. By varying the relative values of $\lambda$ and $\gamma$,  the optimization trades a desire for convexity with a desire for uniformity.  This is reflected in the weight vectors containing negative coefficients and allows a stratification of the data set based on when a negative coefficient emerges. Further, we observe that with $\gamma \ll \lambda \ll 1$, the $\ell_2$-norm of a weight vector is a measure of the distance of the associated point from the boundary of the convex hull which can be used as an alternate means to stratification.

\bibliographystyle{amsplain}
\bibliography{ConvexHullBib}

\end{document}